\documentclass{article}
\usepackage[round,sort&compress,comma,authoryear]{natbib}
\newcommand*{\doi}[1]{\href{http://dx.doi.org/#1}{doi: \nolinkurl{#1}}}
\usepackage[english]{babel}

\usepackage{fullpage}
\usepackage{setspace}
\usepackage{amsmath, amsthm, amssymb}
\usepackage{leftidx}
\usepackage{lscape}
\usepackage{hyperref}
\usepackage{graphicx}
\usepackage{xcolor}
\usepackage{authblk}

\usepackage{array}
\newcolumntype{L}[1]{>{\raggedright\let\newline\\\arraybackslash\hspace{0pt}}m{#1}}
\newcolumntype{C}[1]{>{\centering\let\newline\\\arraybackslash\hspace{0pt}}m{#1}}
\newcolumntype{R}[1]{>{\raggedleft\let\newline\\\arraybackslash\hspace{0pt}}m{#1}}

\title{A coverage criterion for spaced seeds\\ and its applications to SVM string-kernels and $k$-mer distances}
\author[1]{Laurent No\'e }
\author[2]{Donald E. K. Martin }
\affil[1]{{\small ~ LIFL  (UMR 8022 Lille 1/CNRS) - Inria Lille, Villeneuve d'Ascq, France~~~~~~ {\tt laurent.noe@univ-lille1.fr}}}
\affil[2]{{\small ~ Department of Statistics, North Carolina State University, Raleigh, NC, USA~~~~~~~~ {\tt martin@stat.ncsu.edu}}}
\date{\today}

\begin{document}
\maketitle

\begin{abstract}
Spaced seeds have been recently shown to not only detect more alignments, but also to give a more accurate measure of phylogenetic distances~\citep{BodenEtAlGCB13,LeimeisterEtAlBioinformatics14,HorwegeEtAlNAR14},
and to provide a lower misclassification rate when used with Support Vector Machines (SVMs)~\citep{OnoderaShibuyaMLDM13},
We confirm by independent experiments these two results, and propose in this article to use a {\em coverage criterion}~\citep{BensonMakSPIRE08,MartinJSM13,MartinNoeUn14},
to measure the seed efficiency in both cases in order to design better seed patterns. We show first how this {\em coverage criterion} can be directly measured by a full automaton-based approach. We then illustrate how this criterion performs when compared with two other criteria frequently used, namely the {\em single-hit} and  {\em multiple-hit criteria}, through correlation coefficients with the correct classification/the true distance. At the end, for alignment-free distances, we propose an extension by adopting the {\em coverage criterion}, show how it performs, and indicate how it can be efficiently computed. 

\end{abstract}

\section{Introduction}
To detect similarities in bio-sequences, in the so called {\em hit and extend} strategy framework, spaced seeds are now a frequently used technique to define the {\em hit}~\citep{KeichLiMaTrompDAM04}.
Several tools have been proposed that use spaced seeds~\citep{PatternHunter04,HarrisLASTZPhD07,LinZhangZhangMaLiBioinformatics08,BFAST09,ChenSouaiaiaChenBioinformatics09,ZhouMihaiFloreaCIS10,SHRiMP2Bioinformatics11,KielbasaWanSatoHortonFrithGENOMERESEARCH11,BONDBMCBioinformatics13}, or to design spaced seeds~\citep{BuhlerKeichSunJCSS05,KucherovNoeRoytbergJBCB06,IlieEtAlBioinformatics11,AcoSeedANTS12,NuelBTMChapter11,MarshallHermsKaltenbachRahmannTCBB12}.
Work related to spaced seeds also includes the {\em lossless} filtration problem~\citep{BurkhardtKarkkainenFI03,KucherovNoeRoytbergTCBB05,FarachEtAlJCSS07,NicolasRivalsJCSS08,BattagliaEtAlTCS09,GiladiEtAlJCB10,EgidiManziniTCS14,EgidiManziniFI14,BrindaAFL14}, in the sense that all the alignments of a given set must be detected; the work proposed in this article can be applied to this problem too (section~\ref{subsection:computation}), but we concentrate on the {\em lossy} filtration problem in the sense that we suppose that the alignments are associated with a probabilistic model.
We also mention a related work on {\em clump statistics}~\citep{StefanovRobinSchbathDAM07,BassinoClementFayolleNicodemeDMTCS08,MartinColemanJAP11,MarshallHermsKaltenbachRahmannTCBB12,RegnierFangIakovishinaANALCO14} that is close (but not similar), and that can, in some way, be complementary when both of them are considered in a more general framework.

\bigskip

The organization of the article is as follows. Section~\ref{section:notation} gives notation and definitions related to spaced seeds. Section~\ref{section:coverageDefinitionComputation} defines the {\em coverage} of spaced seeds and proposes the tools used to measure it. Section~\ref{section:experiments} shows how {\em coverage} can be used in two biologically oriented applications : first, when spaced seeds are included within SVM kernels (sub-section~\ref{subsection:experimentsSVM}), or when spaced seeds are applied to measure phylogenetic distances (sub-section~\ref{subsection:experimentsDistances}). In this last case, we also propose a new distance based on the coverage (sub-section~\ref{subsubsection:CoverageExperimentalSupport}) and the substantial improvement achieved. Section~\ref{section:concludingRemarks} provides, at the end, some concluding remarks.

\section{Notation}
\label{section:notation}

We suppose here that strings are indexed starting from position number $1$. 
For a given string $u$, we will use the following notation : $u[i]$ gives the $i$-th symbol of $u$, $|u|$ is the length of $u$, and $|u|_a$ is the number of symbol letters $a$ that $u$ contains. Also, ${}_d(u)$ is the prefix of length $d$ of the string $u$, and $(u)_d$ is the suffix of length $d$ of the string $u$. For two strings $u$ and $v$, $u \cdot v$ is the concatenated string. 

Alignments without gaps ({\em indels}) can be modeled by a succession of {\em mismatch} or {\em match} symbols, and thus be represented as a string $x$ in a binary alphabet $\{{\tt 0},{\tt 1}\}$.
A spaced seed can be represented as a string $\pi$, but in a different binary alphabet $\{{\tt *},{\tt 1}\}$ : ${\tt 1}$ indicates a position on the seed $\pi$ where a {\em match} must occur in the alignment $x$ (it is thus called a {\em must match} symbol), whereas $*$ indicates a position where a {\em match} or a {\em mismatch} is allowed (it is thus called a {\em joker} symbol). The {\em weight} of a seed $\pi$ (denoted by $w$) is defined as the number of {\em must match} symbols ($w = |\pi|_1$), whereas the {\em span}/{\em length} of a seed $\pi$ (denoted by $k$) is its full length ($k = |\pi|$).

A spaced seed $\pi$ of length $k$ {\em hits} an alignment $x$ of length $n$ starting at position $i$ ($i \in [1\ldots n-k+1]$) iff 
\[
  \forall j \in [1\ldots k] \qquad \pi[j] = {\tt 1} \implies x[j+i-1] = {\tt 1}
\]

The usual requirement for a seed, when used to detect alignments $x$, is to have {\em at least one hit}~\citep{KeichLiMaTrompDAM04} in $x$, the so called {\em single hit} criterion. Several methods are also based on {\em multiple hits}, as they require more than one hit to trigger an alignment extension~\citep{QuasarRECOMB99,SwiftJCB06,SHRiMP2Bioinformatics11}. In the next section, we extend the way to define criteria based on seed hits by measuring {\em coverage} provided by these hits.

\section{Definition and computation of the seed coverage}
\label{section:coverageDefinitionComputation}
\subsection{Definition of the coverage}
\label{subsection:coverageDefinition}

The {\em coverage} of a seed $\pi$ on an alignment $x$ is defined by the number of ${\tt 1}$'s in the alignment $x$ that are covered by {\em at least} one {\em must match} symbol of one of the seed's hits~\citep{BensonMakSPIRE08,MartinJSM13,MartinNoeUn14}.

For example, the seed $\pi = {\tt 1 1 * 1}$ has three hits on the string alignment $x = {\tt 1 0 1 1 1 1 0 0 1 0 1 1 1 1 1}$.
The coverage provided by these hits (denoted by $\bullet$ symbols below) is $8$.

\begin{center}
\begin{tabular}{c|cccccccccccccccc}
$\pi \; {}_{occ_1}$ &  &   & 1 & 1 & * & 1 &   &   &   &   &   &  &   &   &  \\[-3.2mm]
$\pi \; {}_{occ_2}$ &  &   & \vdots & \vdots &   & \vdots &   &  &   &   & 1 & 1 & * & 1 &  \\[-3.2mm]
$\pi \; {}_{occ_3}$ &  &   & \vdots & \vdots &   & \vdots &   &  &   &   & \vdots & 1 & 1 & * & 1\\[-0.4mm]
\hline
$x$ & {\tt 1} & {\tt 0} &  $\underset{\bullet}{\tt 1}$ & $\underset{\bullet}{\tt 1}$& {\tt 1} &  $\underset{\bullet}{\tt 1}$ & {\tt 0} & {\tt 0} & {\tt 1} & {\tt 0} &  $\underset{\bullet}{\tt 1}$ &  $\underset{\bullet}{\tt 1}$ &  $\underset{\bullet}{\tt 1}$ &  $\underset{\bullet}{\tt 1}$ &  $\underset{\bullet}{\tt 1}$\\[1mm]
\end{tabular}
\end{center}

The coverage concept can be generalized to multiple seed patterns. For example, the set of seeds $\{\pi_1,\pi_2\} = \{ {\tt 1 1 * 1} , {\tt 1 * 1 * 1} \}$ has six hits on the string alignment $x$. The coverage provided by these hits is $11$.

\begin{center}
\begin{tabular}{c|ccccccccccccccc}
$\pi_{2}  \; {}_{occ_1}$ &  1 &  *  & 1  & *  & 1 \\[-3.2mm]
$\pi_{1}  \; {}_{occ_2}$ & \vdots &   & 1 & 1 & * & 1 \\[-3.2mm]
$\pi_{2}  \; {}_{occ_3}$ & \vdots &   & \vdots & \vdots & \vdots & \vdots &   &  & 1 & * & 1 & * &  1 \\[-3.2mm]
$\pi_{1}  \; {}_{occ_4}$ & \vdots &   & \vdots & \vdots & \vdots & \vdots &   &  & \vdots &   & 1 & 1 & * & 1 &  \\[-3.2mm]
$\pi_{2}  \; {}_{occ_5}$ & \vdots &   & \vdots & \vdots & \vdots & \vdots &   &  & \vdots &      & 1 & * & 1 & * &  1\\[-3.2mm]
$\pi_{1}  \; {}_{occ_6}$ & \vdots &   & \vdots & \vdots & \vdots & \vdots &   &  & \vdots &   & \vdots & 1 & 1 & * & 1\\[-.4mm]

\hline
$x$ &$\underset{\bullet}{\tt 1}$ & {\tt 0} & $\underset{\bullet}{\tt 1}$ & $\underset{\bullet}{\tt 1}$ &$\underset{\bullet}{\tt 1}$ & $\underset{\bullet}{\tt 1}$ & {\tt 0} & {\tt 0} & $\underset{\bullet}{\tt 1}$ & {\tt 0} & $\underset{\bullet}{\tt 1}$ & $\underset{\bullet}{\tt 1}$ & $\underset{\bullet}{\tt 1}$ & $\underset{\bullet}{\tt 1}$ & $\underset{\bullet}{\tt 1}$\\[1mm]
\end{tabular}
\end{center}

\subsection{Coverage automaton}
\label{subsection:coverageAutomaton}
Given a seed $\pi$ or a set of seeds $\{\pi_1,\pi_2,\ldots\}$ along with an input string $x$, the aim of the automaton is to compute the coverage of $\pi_1,\pi_2,\ldots$ on $x$, as defined in section~\ref{subsection:coverageDefinition}. 
To fully compute the coverage, a necessary and sufficient task, typically devoted to an automaton, is to update the coverage each time we concatenate a new symbol to the right of $x$.
For example, for the set of seeds $\{\pi_1,\pi_2\} = \{ {\tt 1 1 * 1} , {\tt 1 * 1 * 1} \}$ and the string $x = {\tt 1 0 1 1 1 1 0 0 1 1 1 1 0}$, we desire to determine the set of newly covered positions (denoted by two $\circ$ symbols below) after reading the new symbol ${\color{gray}{1}}$ to form the extended string $x' = x \cdot {\color{gray}{1}}$, together with their count to update the coverage. We will call this (+2) value the {\it coverage increment}.

\begin{minipage}[h]{\textwidth}
~\\
\begin{center}
\begin{tabular}{c|cccccccccccccc}
$\pi_{2}  \; {}_{occ_1}$ &  1  &  *  & 1  & *  & 1  \\[-3.2mm]
$\pi_{1}  \; {}_{occ_2}$ &  \vdots &     & 1 & 1 & * & 1  \\[-3.2mm]
$\pi_{1}  \; {}_{occ_3}$ &  \vdots &     & \vdots & \vdots & \vdots & \vdots &   &   & 1 & 1 & *  & 1 \\[-0.4mm]
\hline
$x$ &$\underset{\bullet}{\tt 1}$ & {\tt 0} & $\underset{\bullet}{\tt 1}$ & $\underset{\bullet}{\tt 1}$ &$\underset{\bullet}{\tt 1}$ & $\underset{\bullet}{\tt 1}$ & {\tt 0} & {\tt 0} & $\underset{\bullet}{\tt 1}$ & $\underset{\bullet}{\tt 1}$ & {\tt 1} & $\underset{\bullet}{ \tt 1}$ & {\tt 0} & {\tt ~ }\\[1mm]
\end{tabular}\\
~\\
$\boldsymbol{\downarrow}$
~\\
\begin{tabular}{c|ccccccccccccccc}
$\pi_{2}  \; {}_{occ_1}$ &  1  &  *  & 1  & *  & 1  \\[-3.2mm]
$\pi_{1}  \; {}_{occ_2}$ &  \vdots &     & 1 & 1 & * & 1  \\[-3.2mm]
$\pi_{1}  \; {}_{occ_3}$ &  \vdots &     & \vdots & \vdots & \vdots & \vdots &   &   & 1 & 1 & *  & 1 \\[-3.2mm]
$\pi_{2}  \; {}_{occ_4}$ &  \vdots &     & \vdots & \vdots & \vdots & \vdots &   &   & \vdots & 1  & *  & 1 & * & {\color{gray}{1}} \\[-3.2mm]
$\pi_{1}  \; {}_{occ_5}$ &  \vdots &     & \vdots & \vdots & \vdots & \vdots &   &   & \vdots & \vdots & 1 & 1 & * & {\color{gray}{1}} \\[-0.4mm]
\hline
$x'$ &$\underset{\bullet}{\tt 1}$ & {\tt 0} & $\underset{\bullet}{\tt 1}$ & $\underset{\bullet}{\tt 1}$ &$\underset{\bullet}{\tt 1}$ & $\underset{\bullet}{\tt 1}$ & {\tt 0} & {\tt 0} & $\underset{\bullet}{\tt 1}$ & $\underset{\bullet}{\tt 1}$ & $\underset{\circ}{\tt 1}$ & $\underset{\bullet}{ \tt 1}$ & {\tt 0} & $\underset{\circ}{{\color{gray}{\tt 1}}}$\\[1mm]
\end{tabular}
\end{center}
~\\
\end{minipage}

For a set of seeds $\{\pi_1,\pi_2,\ldots\}$ with $k = max_i (|\pi_i|)$, first notice that a suffix of $x$ of length (at most) $k-1$ is sufficient to know which {\em proper prefixes} of one of the seeds can lead to a new hit : we will call $q$ this suffix. Moreover, to update the coverage increment, we need to know which {\tt 1} symbols inside $q$ have already been covered by previous hits of one of the seeds : this can be done with a binary word $c$ of length $|q|$ associated with $q$. States of the automaton are thus defined accordingly by a pair $\langle\substack{q^{} \\ c^{}}\rangle$.

For example, for the set of seeds $\{\pi_1,\pi_2\} = \{ {\tt 1 1 * 1} , {\tt 1 * 1 * 1} \}$, the state reached when reading the first string alignment $x = {\tt 1 0 1 1 1 1 0 0 1 1 1 1 0}$ (used in the previous example) is represented by the pair 
$
\langle\substack{q^{} \\ c^{}}\rangle 
= 
\langle\substack{\tt 1 \\ \bullet}\substack{\tt 1\\~}\substack{\tt 1\\\bullet}\substack{\tt 0\\~}\rangle
$
and the transition to $x' = x \cdot {\color{gray}{1}}$ can be computed accordingly with the new hits of $\pi_1$ and $\pi_2$
\begin{center}
\begin{tabular}{c|cccccc}
${\color{gray}{\pi_2 \; {}_{occ_4}}}$ & \ldots & 1                           & *                                       & 1                           & *       & {\color{gray}{1}}                         \\[-3.2mm]
${\color{gray}{\pi_1 \; {}_{occ_5}}}$ & \ldots &                             & 1                                       & 1                           & *       & {\color{gray}{1}}                         \\[-0.4mm]
\hline
$q \rightarrow {\color{gray}{q'}}$                               & \ldots & $\underset{\bullet}{\tt 1}$ & $\underset{\color{gray}{\circ}}{\tt 1}$ & $\underset{\bullet}{\tt 1}$ & {\tt 0} & {\color{gray}{$\underset{\circ}{\tt 1}$}} \\[1mm]
\end{tabular}
\end{center}

\begin{landscape}
\begin{figure}
 \centering
  \caption[Mealy coverage automaton (count on transitions)]{\label{fig:MealyCoverageAutomaton}Minimized Mealy coverage automaton (count on transitions) for the seeds $\{\pi_1,\pi_2\} = \{ {\tt 1 1 * 1} , {\tt 1 * 1 * 1} \}$}
  \includegraphics[width=1.2\textwidth]{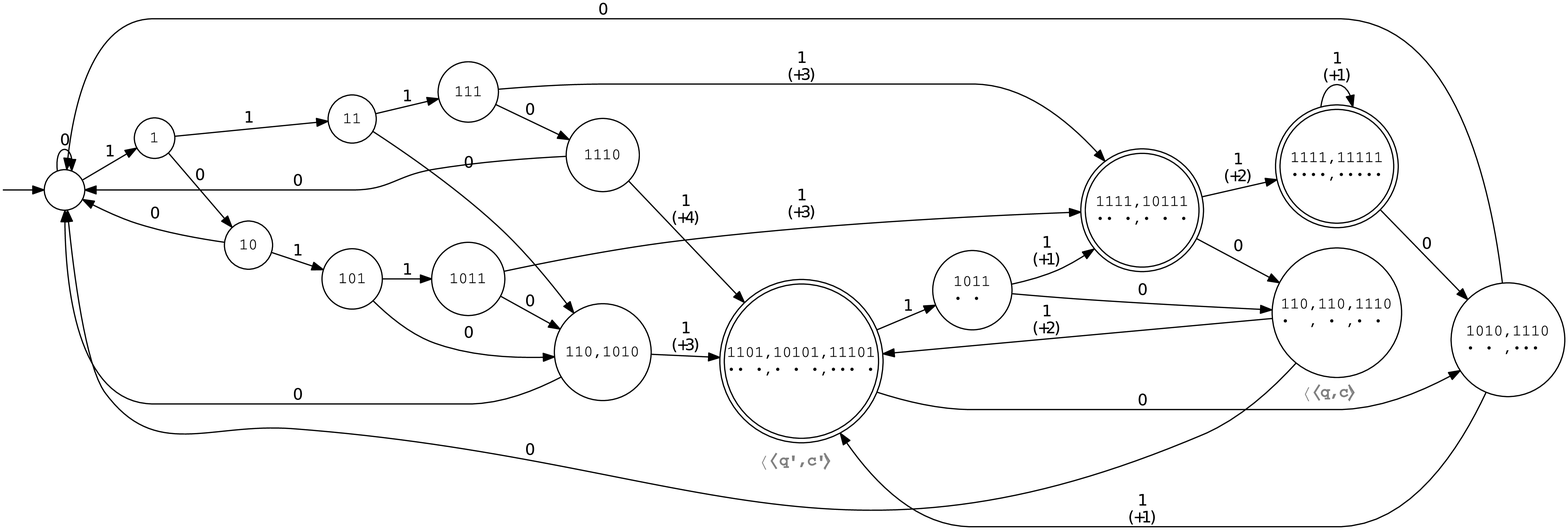}
\end{figure}
\begin{figure}
 \centering
  \caption[Moore coverage automaton (count on states)]{\label{fig:MooreCoverageAutomaton}Minimized Moore coverage automaton (count on states) for the seeds $\{\pi_1,\pi_2\} = \{ {\tt 1 1 * 1} , {\tt 1 * 1 * 1} \}$}
  \includegraphics[width=1.2\textwidth]{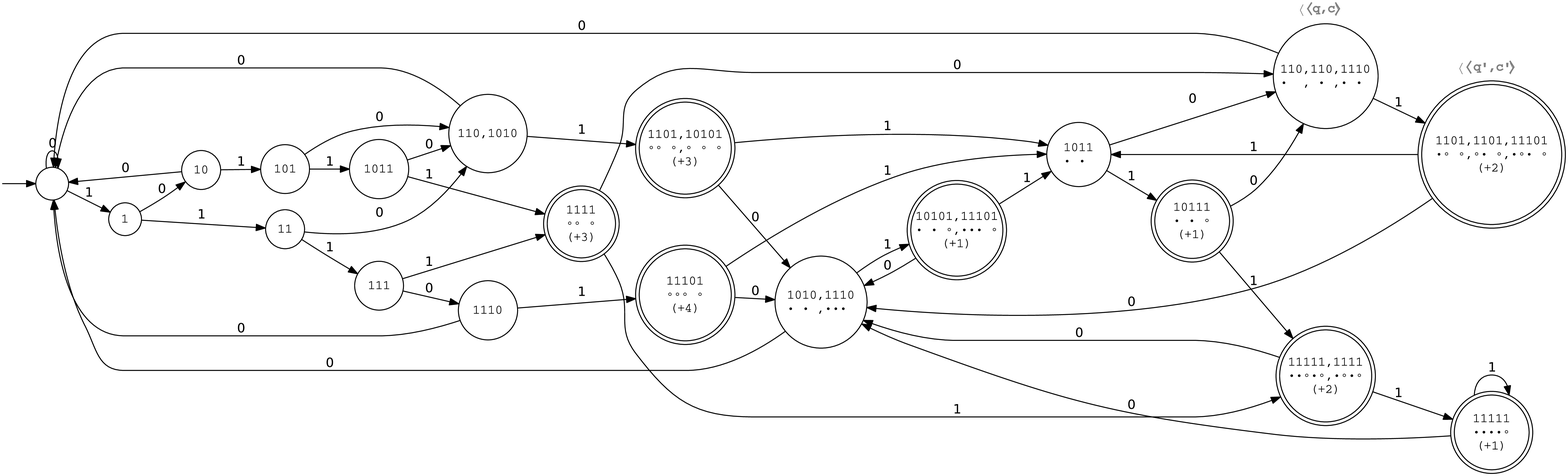}
\end{figure}
\end{landscape}

The new state may be represented by the pair 
$
\langle\substack{q' \\ c'}\rangle = \langle\substack{\tt 1\\\bullet}\substack{\tt 1\\\bullet}\substack{\tt 0\\~}\substack{\tt 1\\\bullet}\rangle
$
with $|q'|\leq k-1$.
Note that $q'$ can even be reduced to a smaller suffix, because no {\em proper prefix} of $\pi_1$ or $\pi_2$ can start with $q' = {\tt 1101}$, but a prefix of $\pi_2$ can match the first proper suffix of $q'$, namely {\tt 101}, to initiate a hit. Thus $\langle\substack{q' \\ c'}\rangle = \langle\substack{\tt 1\\\bullet}\substack{\tt 1\\\bullet}\substack{\tt 0\\~}\substack{\tt 1\\\bullet}\rangle \equiv \langle\substack{\tt 1\\\bullet}\substack{\tt 0\\~}\substack{\tt 1\\\bullet}\rangle$
 : this reduction can be done easily using the Fail function of Aho-Corasick algorithm~\citep{AhoCorasick74} which is applied in classical seed automata~\citep{BuhlerKeichSunJCSS05,KucherovNoeRoytbergCIAA07}, as well as coverage automata~\citep{BensonMakSPIRE08,MartinNoeUn14}. We will suppose that we always apply this reduction on all the states $\langle\substack{q \\ c}\rangle$.

\medskip

From the point of view of the automaton definition, two finite state machines are possible : {\em Mealy } or {\em Moore}. Accordingly, the automaton must provide the {\em coverage increment}, either on each transition (for the {\em Mealy} automaton), or on each state (for the {\em Moore} automaton). For example, on the set of seeds $\{\pi_1,\pi_2\} = \{ {\tt 1 1 * 1} , {\tt 1 * 1 * 1} \}$, these two representations are illustrated on Figures~\ref{fig:MealyCoverageAutomaton} and~\ref{fig:MooreCoverageAutomaton} : due to size, we present here the minimal version for both automata by merging equivalent states. For readability, when some hits occur, we have represented the states $\langle\substack{q \\ c}\rangle$ with their full length matching symbols of length up to $k$ and not $k-1$ (see for example $\langle\substack{q^{} \\ c^{}}\rangle \equiv \langle\substack{\tt 1\\\bullet}\substack{\tt 1\\\bullet}\substack{\tt 0\\~}\substack{\tt 1\\\bullet}\rangle $ and $\langle\substack{q' \\ c'}\rangle \equiv \langle\substack{\tt 1\\\bullet}\substack{\tt 0\\~}\substack{\tt 1\\\bullet}\rangle$ on the Figures~\ref{fig:MealyCoverageAutomaton} and~\ref{fig:MooreCoverageAutomaton}). 

The {\em Mealy} automaton is obviously more compact when considering the number of states. On the other hand, it requires one to store an additional value per transition (and also needs more specific algorithms : for example, the Hopcroft minimization algorithm~\citep{Hopcroft71} must be adapted to the Mealy case).

Each representation has been independently implemented by one of the authors : the one based on {\em count on transition} ({\em Mealy}) is implemented in Matlab~\citep[see][]{MartinJSM13,MartinNoeUn14},  and code has been also tested on Octave~\citep{Octave14}, whereas the other, mainly for compatibility issues, is based on {\em count on states} ({\em Moore}), and is generalized for subset seeds (slight extension of spaced seeds) with multiple seeds in mind~\citep{KucherovNoeRoytbergCIAA07}. The ``Mealy'' Matlab code is available upon request from the second author, and the ``Moore'' code is included in the C++ Iedera program~\citep{IederaSoftwareUnpublished} starting from development version 1.06 $\alpha7$.

Several minimizations of the states (considering both $q$ and $c$) can be considered {\em during} the construction of these automata, but the details are out of scope of this article \citep[see][]{KucherovNoeRoytbergCIAA07,MartinNoeUn14}. In practice, we use at least two methods to detect coverage strings $c$ that are {\em equivalent}, together with the optimisation of~\cite{KucherovNoeRoytbergCIAA07} on strings $q$ to save some memory space {\bf before} completing the full automaton. Note that this last automaton, once entirely built, can always be fully reduced to its minimal form, for example by applying the classical Hopcroft minimization algorithm~\citep{Hopcroft71}.

Independently, we also mention that it seems difficult, for this special {\em coverage} problem, to find an equivalent classical {\em regular expression} to help build the automata. Even classical tools (such as {\tt grep}) have for example equivalent parameters to simulate a {\em single} or {\em multiple} hit, but no parameter is provided for this {\em coverage} problem.

\subsection{Computation}
\label{subsection:computation}
Given a generative model $\mathcal{X}$ for the string $x$, it is possible to compute the distribution of the coverage values according to a Markov process~\citep{MartinJSM13,MartinNoeUn14} or any model that can be represented by a non-deterministic probabilistic automaton~\citep{KucherovNoeRoytbergJBCB06,NuelJAP08,MarshallHermsKaltenbachRahmannTCBB12,MartinNoeUn14}. We don't consider directly in this article this computation, as the model used in our tests is pure Bernoulli : the computation can thus be performed directly with a simple dynamic programming algorithm on the coverage automaton of section~\ref{subsection:coverageAutomaton}. We refer to the aforementioned articles for more details on more complex probabilistic models.

Independently, we also mention that the work proposed here is applied on the {\em lossy seed framework}, in the sense that we consider a {\em probability} to {\em hit} (or {\em cover}) an alignment sequence $x$ generated by a model $\mathcal{X}$. However, this work is not strictly limited to probabilities, and can be easily extended, for example to the {\em lossless seed framework}. In that case, the set of alignments is fixed, for example by giving a fixed length together with a fixed number of errors : the problem is then to always {\em hit} (or {\em cover}) any of the alignments on this set (so without {\em loss}). This last computation can be done easily, simply by replacing the {\em semi-ring} used for probabilities by a less conventional {\em tropical} semi-ring~\citep{Simon88,Pin98,MohriHWAChapter09} used for match/mismatch scores or mismatch costs. Note also that the simple fact of counting the number of alignments, in alignment classes that have a given percentage of identity \citep[as done in][]{BensonMakSPIRE08}, or a given coverage for a set of seeds, or any combination of these elements, is also possible, by use of a {\em counting} semi-ring adapted for this task~\citep{HuangDPRing06}.

\section{Experiments}
\label{section:experiments}

In this section, we consider two {\em biological sequence} oriented applications that have recently been proposed to use spaced seeds : SVM classifiers based on spaced string kernels~\citep{OnoderaShibuyaMLDM13}, and alignment-free distance estimators using spaced $k$-mers~\citep{BodenEtAlGCB13,LeimeisterEtAlBioinformatics14,HorwegeEtAlNAR14}. We show that the coverage sensitivity can be used in both cases to improve the estimators, and thus also be applied to the selection of the best seed patterns on such domains.

Additional data and results, together with scripts used for this section can be found at\newline \url{http://bioinfo.lifl.fr/yass/iedera_coverage/}.

\subsection{Coverage sensitivity and spaced seed string kernels}
\label{subsection:experimentsSVM}
String kernels~\citep{LodhiEtAlJMLR02} are a classical model used for text classification with SVM. They have frequently been applied to biological sequence classification, as $k$-spectrum kernels~\citep{LeslieEtAlPSB02}, mismatch $k$-spectrum kernels~\citep{LeslieEtAlBioinformatics04}, string alignment kernels~\citep{SaigoEtAlBioinformatics04}, profile-based string kernels~\citep{KuangEtAlJBCB05} to cite a few examples.

$k$-spectrum kernels and its derivatives are mostly used with {\em contiguous seeds} : surprisingly, no {\em spaced seeds} were designed to comply with this approach. However, it has been experimentally shown by~\cite{OnoderaShibuyaMLDM13}, and during the submission of this work by~\cite{GhandiEtAlPLoSComputationalBiology14,GhandiEtAlJMB14} that spaced seeds help decrease the zero/one misclassification rate in practice, even for the simplest kernels.
The main reason of this lack might be the intrinsic difficulty to find a {\em correct estimation criterion} for spaced seed patterns, but on the other hand, not so much effort has been made to increase the diversity of criteria used. Most of the proposed algorithms to estimate spaced seed sensitivity concentrate on the {\bf single-hit criterion} ({\em ``at least one hit for a seed/set of seeds''}). This criterion makes sense for classical ``hit and extend'' alignment methods used in bioinformatics, but seems to be too restrictive for spectrum kernels that are supposed to filter the information content based of ``several concordant clues''.

The {\bf multi-hit criterion} ({\em ``at least $t$ hits for a seed/set of seeds''}) seems at first more appropriate for this task, but again has never been tried in this field of research. One possible drawback is that it does not distinguish highly overlapped hits of seeds from disjoint ones. Finally, and for the latter reason, we also decided to apply the new {\bf coverage criterion} ({\em ``at least $t$ covered {\tt 1}-symbols in the alignment, each covered by at least one {\tt 1}-symbol of a seed hit ''}) in comparison with the two others.

In the two following subsections, we try to correlate these three criteria with SVM zero/one misclassification rate.

\subsubsection{SVM-Benchmark and Protocol}

The benchmark used for this test consists of 2208 families extracted from the non-coding RNA database RFAM v11.0~\citep{RFAM11}. It represents up to 65908 sequences per family.

We decided to split each family by randomly picking 50\% of its sequences for the SVM learning process and keeping the 50\% remaining for the classifier to measure the zero/one misclassification rate. We use the $SVM^{multiclass}$~\citep{Thorsten02} package {\small Version 2.20 (date 14.08.2008)} from {\small\url{http://www.cs.cornell.edu/people/tj/svm_light/svm_multiclass.html}} with the linear kernel.
In each case, single or double seeds of weight 3 and span from 3 to 7 were used as a $k$-spectrum string kernel.

\subsubsection{Seed sensitivity}

In parallel, for each single or double seed, we compute its ``sensitivity'', either using the single hit criterion, the multi-hit criterion, or the coverage criterion. Note that, for the two last criteria, we have the possibility to change the {\em threshold} $t$ required to consider a success.
We arbitrarily choose to measure these seeds on an i.i.d. alignment model of length 32 (the probability to have a 1-symbol in the alignment has been fixed at $0.7$) although experiments show that this does not have much influence on the final results [data not shown].

\begin{figure}[htb]
 \centering
 \caption[]{\label{fig:zerooneVSsensitivity} {\em zero/one misclassification rate} vs {\em theoretical sensitivity}}
 \includegraphics[width=.5\textwidth]{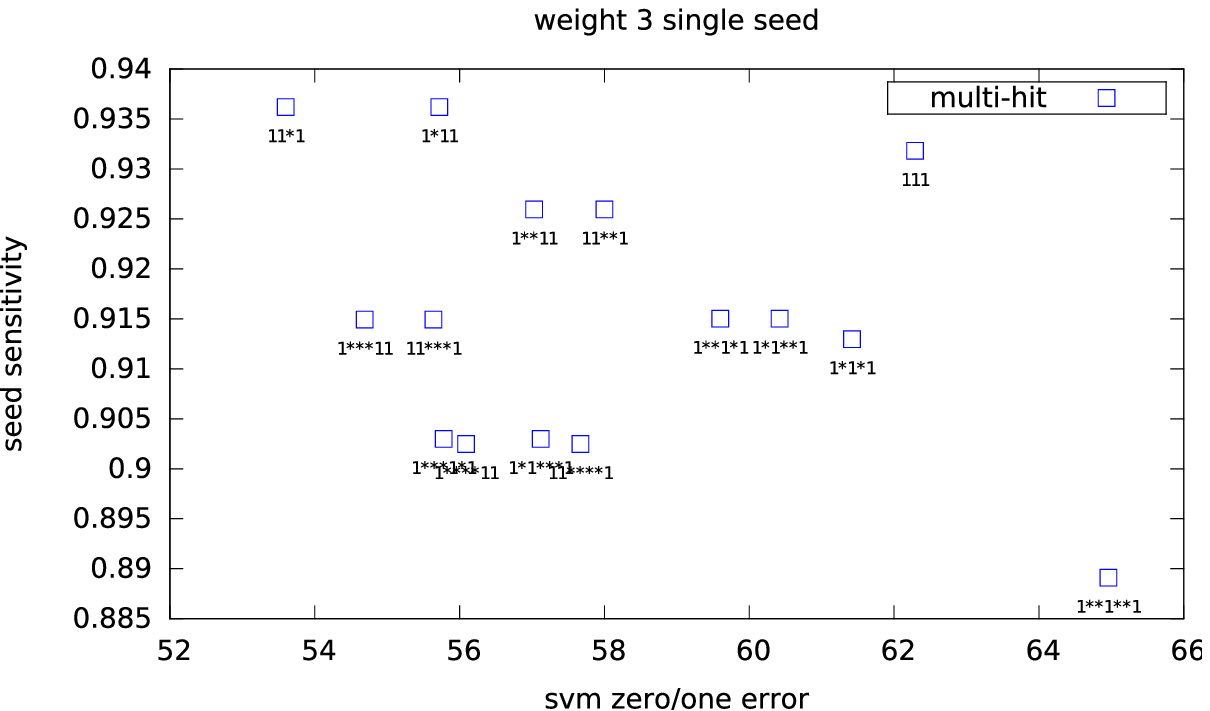}~ \includegraphics[width=.5\textwidth]{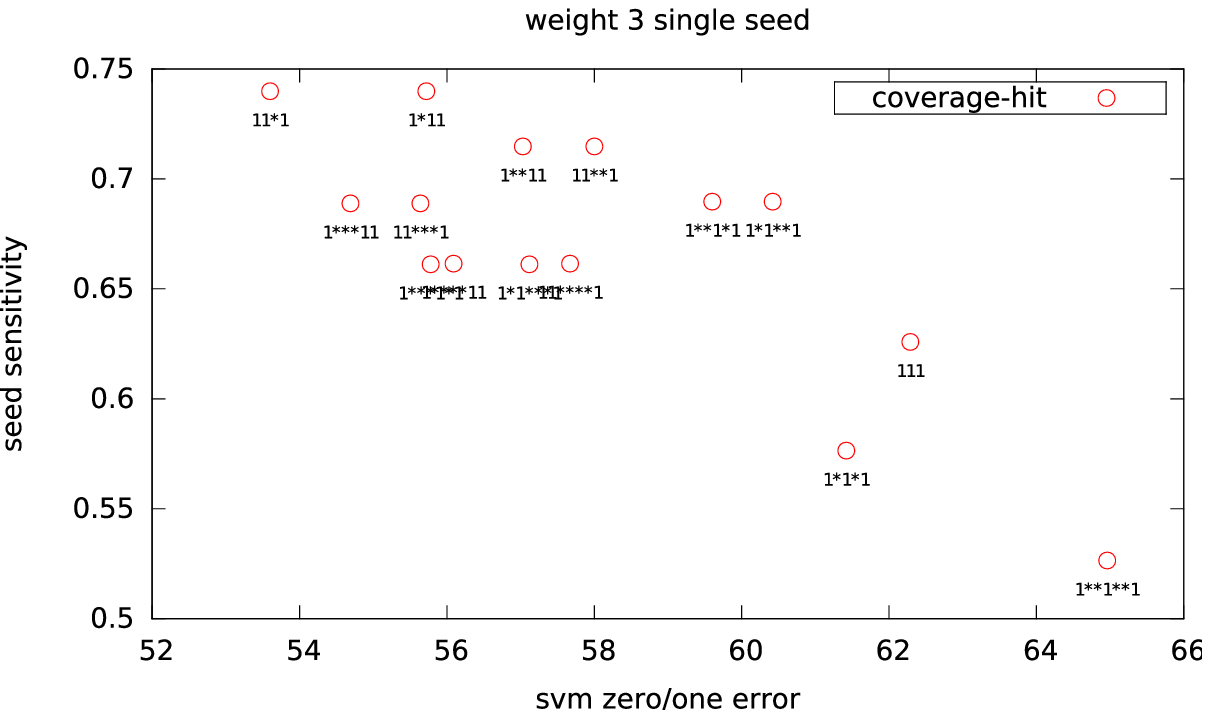}
\end{figure}

Examples of comparative plots are given in Figure \ref{fig:zerooneVSsensitivity} for multi-hit and coverage-hit : a slight correlation can be seen at first sight. 
But we can also see that some seeds with repetitive and highly correlated patterns (e.g. {\tt\small 1*1*1}), usually bad in theory, are in practice more efficient for the SVM-classifier.

\subsubsection{Correlation between zero/one misclassification and the three criteria}

Finally, to determine if one of the three estimators was better suited to correlate with this SVM classifier task, we computed the {\em sample Pearson correlation coefficient} for each of the three criteria, for each set of seeds : this gives the best correlation between the theoretical seed sensitivity estimated by one of the three estimators with the experimentally measured sensitivity of the SVM classifier of each set of seeds.

\begin{figure}[htb]
 \centering
 \caption[]{\label{fig:zerooneCORsensitivity}Correlation coefficient between {\em zero/one misclassification} rate and {\em theoretical sensitivity}}
\includegraphics[width=.5\textwidth]{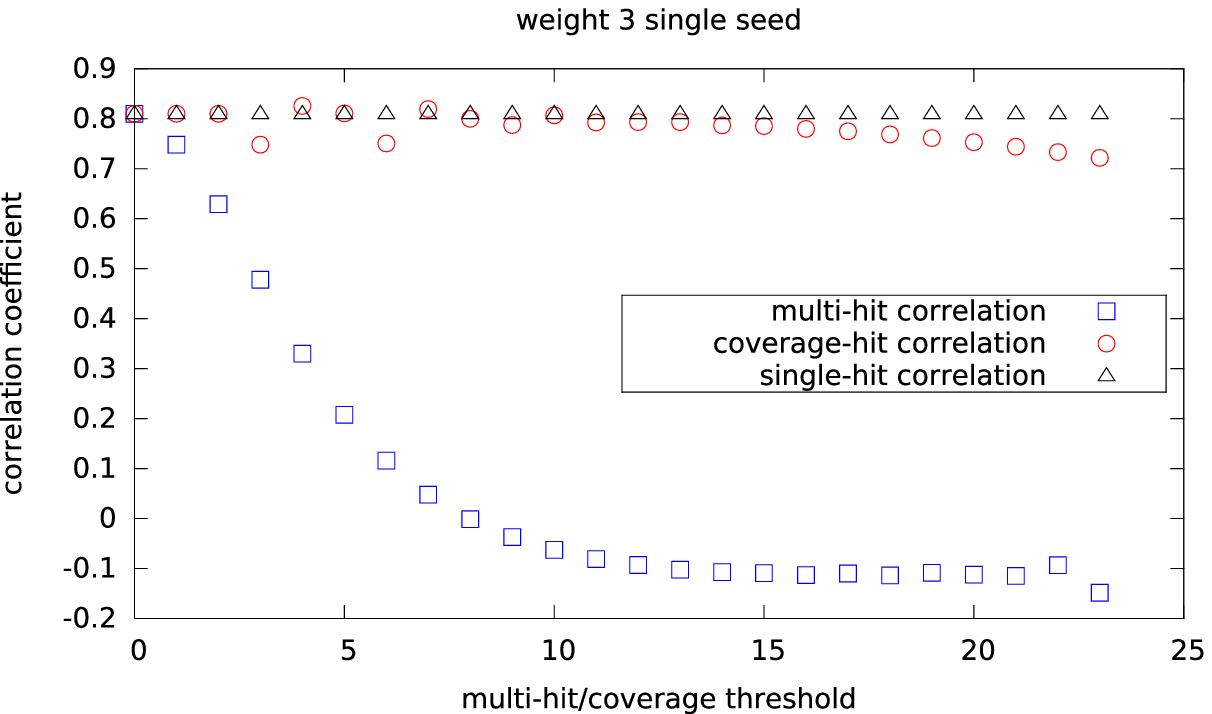}~\includegraphics[width=.5\textwidth]{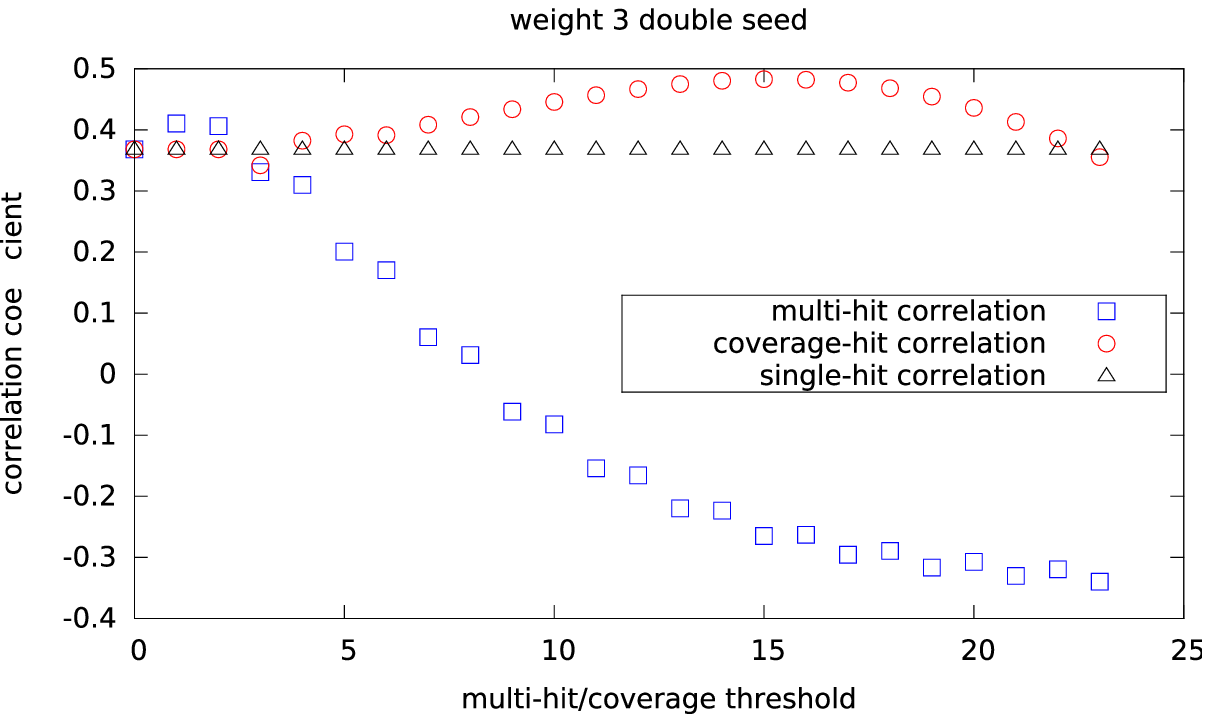}
\end{figure}

For both the multiple hit and coverage criteria, we allowed the threshold parameter $t$ for seed sensitivity to vary ($x$-axis). These results are illustrated in Figure \ref{fig:zerooneCORsensitivity} for single and double seeds.

Surprisingly, correlation results for the {\bf multi-hit criterion} are not good when the number of hits required is too large. This must be taken into account when using this criterion because the multi-hit criterion gives correct results for double seeds when the number of hits is for example at 2 . 

The {\bf single-hit criterion} gives good results for each set. However combining single and double seeds into one set, and doing the same experiment makes it the worse of the three estimators [data not shown].
A carefully chosen value for the {\bf coverage criterion} (here between 14 and 16) helps to reach the highest correlation of the three for double seeds. On single seeds, this is difficult to conclude, due to the few seeds of weight 3 that have been tested. Note that we also tried the same experiment for seeds of weight 4 but the dimension used here ($4^4$) makes the classifier more random without a preselection of dimensions [data not shown].

We can first notice that the correlation of the single-hit criterion is more stable than the correlation of the coverage criterion that varies more for lower thresholds. It also seems that the optimal coverage threshold is at some point a surprisingly quite regular and convex function that might be estimated when enough data is available.

\subsection{Coverage sensitivity and alignment-free distance for sequence comparison}
\label{subsection:experimentsDistances}

Estimating alignment-free distance is a common method used for sequence comparison in multiple alignment tools guided tree estimation~\citep{Muscle04} and related phylogenetic tree estimation~\citep{CVTree04,LiuEtALNAR08}.
Several distances are based on fixed size $k$-mers~\citep{VingaAlmeidaBioinformatics03,SimsaEtALPNAS09} with possible mismatches allowed~\citep{ApostolicoEtAlDCC14}, or with variable length $k$-mers : local decoding~\citep{DidierEtAlTCS12}, irredundant common subwords~\citep{CominVerzottoAMB12}, etc. They are applied on assembled genomes~\citep{HauboldEtAlBMCBioinformatics05,ChorEtAlGenomeBiology09}, protein classification~\citep{StropeaMoriyamaGenomics07}, and even on unassembled genomic data to estimate phylogenies~\citep{MaurerStrohEtAlJBCB13}. We refer to a recent special issue on alignment-free methods for more details~\citep{VingaBriefingsinBioinformatics14}.

Interestingly, it's only in the last year that the use of spaced seeds has been proposed~\citep{BodenEtAlGCB13,LeimeisterEtAlBioinformatics14,HorwegeEtAlNAR14}, with recent applications for specific {\em Next Generation Sequencing} tasks, such as multi-clonal clusterization~\citep{VidjilBMCGenomics14}. Here again, the lack of seed criteria used in the literature didn't help the selection of good seeds for these tasks.

In subsection~\ref{subsubsection:MultihitExperimentalSupport}, we recall that the ``classical'' distance can be estimated by multi-hit sensitivity computation which helps in selecting good spaced seeds.

In subsections~\ref{subsubsection:CoverageExperimentalSupport} and~\ref{subsubsection:CoverageAlgorithm}, we also show that coverage sensitivity can be used in a more elaborate distance : this distance can be computed using the {\em Longest Increasing Subsequence (LIS)} of the positions of the common hits between gapped $k$-mers. As LIS can be computed in $t \cdot \log (t)$ time, where $t$ is the number of hits, it is thus a reasonable estimator in practice.

\subsubsection{Multi-hit experimental support}
\label{subsubsection:MultihitExperimentalSupport}

One common method used to estimate alignment-free distances is based on $k$-mer frequency : $4^k$ counts can be first made and used as simple {\em Feature Frequency Profiles} (where counts are normalized to relative frequencies for any of the $4^k$ $k$-mers), or more elaborate {\em Composition Vectors} (where normalization is done with the help of a background model). Distances can then be estimated by several models~\citep{VingaAlmeidaBioinformatics03} to provide phylogenetic applications with an initial distance matrix. As some of these phylogenetic methods, as {\em Unweighted Pair Group Method with Arithmetic Mean} (UPGMA)~\citep{MichenerSokalEvolution57}, start by considering small distances, it's important to have the best estimator here, and keep track of common $k$-mers (or spaced $k$-mers) and their common locations in the two sequences. One estimator that can help in that task is the number of seed hits obtained : we will call it the {\em multi-hit value}.

For our experiment, we use a set of seeds (627 seeds of weight 3 or 4, span up to 7, single seed or double seed patterns), a percentage of identity varying from 20\% to 100\% by steps of 5\% each time, and we generate (for each percentage of identity) 1000 alignments of length 32. We then measure the {\em multi-hit value} of each alignment and compare it to the true alignment distance.

It can be shown first (Figure \ref{fig:distanceEstimatorCorrelation} $x$-axis only) that the correlation coefficient is high ($>0.9$ for seeds of weight 3, less otherwise). Provided that we expect to pay a little additional cost, it is possible to improve this result, as shown in the next section.

\subsubsection{Coverage experimental support}
\label{subsubsection:CoverageExperimentalSupport}
The distance we propose to measure is based on the number of covered {\tt 1}-symbols in the alignment, each covered by at least one {\tt 1}-symbol of a seed hit : we will call it the {\em coverage value}.
To show how this distance better estimates the true distance (we assume here that the Hamming distance is the true distance), we are repeating the same experiment with both the {\em multi-hit value} and the {\em coverage value} on the set.

We use the same protocol here : the same set of seeds (627 seeds of weight 3 or 4, span up to 7, single seed or double seed patterns), the same percentage of identity varying from 20\% to 100\% by steps of 5\% each time, and generating for each percentage of identity the same 1000 alignments of length 32 each time, we measure the {\em multi-hit {\bf and} the coverage values} for each simulated alignment. Then, we compared the {\em correlation coefficient} for each of these two measures with the true percentage of identity used to simulate the alignment.

\begin{landscape}
\begin{figure}[htb]
 \centering
 \caption[]{\label{fig:distanceEstimatorCorrelation}}
 \includegraphics[width=1.25\textwidth]{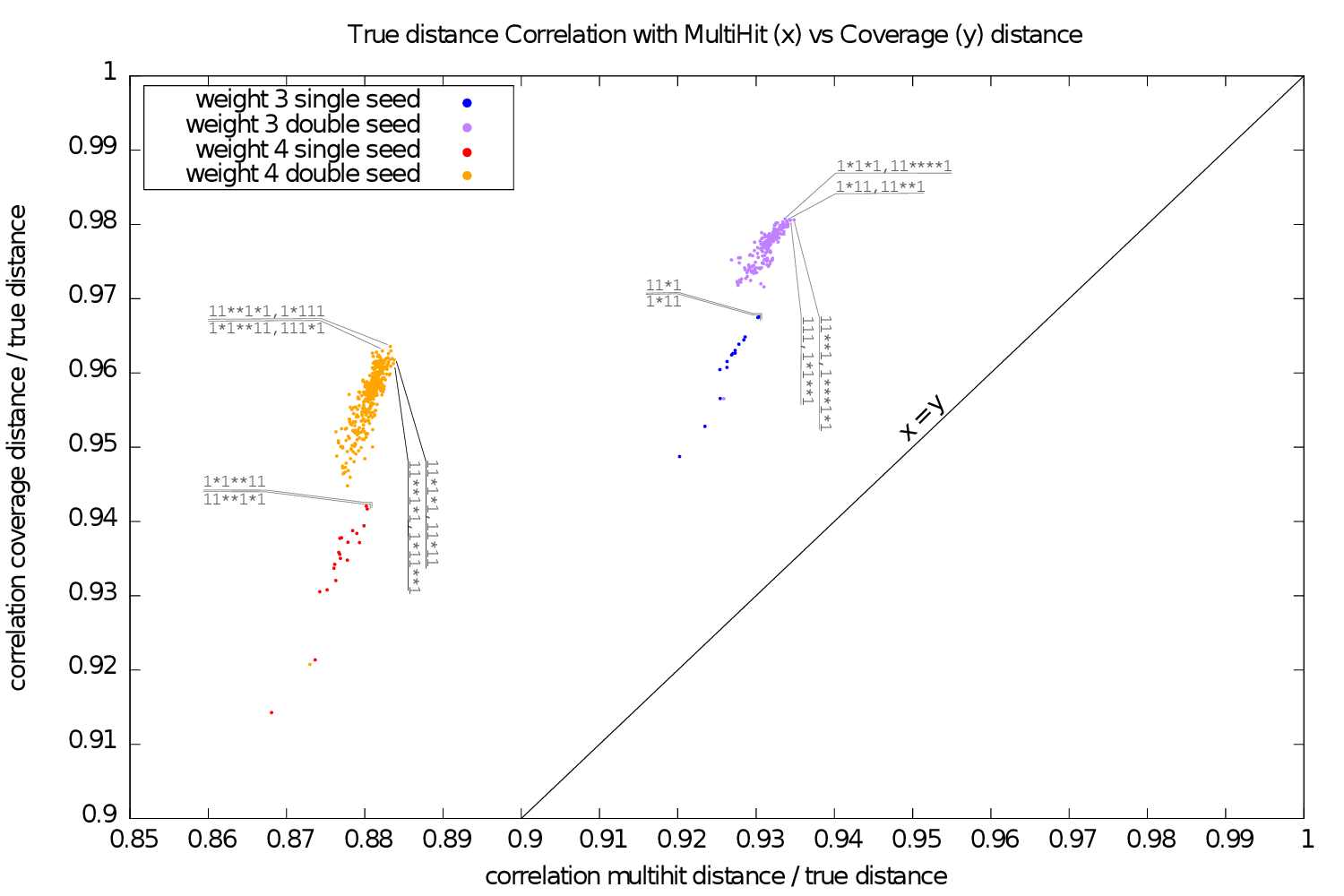}
\end{figure}
\end{landscape}

The {\em correlation coefficient} for all the seeds was 0.88 for the {\em multi-hit value} and 0.96 for the {\em coverage value}. We tried to refine this first experiment by separately measuring single seed patterns and double seed patterns and running the same test. For single seed patterns, the correlation was 0.89 and 0.94 respectively, whereas for multiple seed patterns, it was 0.89 and 0.96 respectively. We also tried to measure this correlation for each of the 627 seeds : Figure \ref{fig:distanceEstimatorCorrelation} plots these two correlations (pair of coordinates).

Note first that, as all the points for this plot are on the {\em left-upper} region, the true distance is better estimated by the {\em coverage value} than by the {\em multi-hit value}. We can also notice that double seeds outperform single seeds in both cases, so that multiple seed patterns can help in estimating the distance more accurately than single seed patterns : the gain is even better for the {\em coverage value} than for the {\em multi-hit value}.

From the point of view of the seed weight and the number of seed patterns used, we can see that using {\em two patterns of weight 4} gives the same correlation coefficient as using {\em one single pattern of weight 3}, {\bf but only for the coverage value, not for the multi-hit value} : this encouraging result may help defend the idea that more patterns of larger weight will help measure a correct distance. Note that this conclusion is quite similar to the one provided ten years ago for detecting alignments~\citep{PatternHunter04}, which was recently and independently observed in~\cite{LeimeisterEtAlBioinformatics14,HorwegeEtAlNAR14}, but here, as the distance estimation problem is quite different from alignment detection, the seeds designed will probably be completely different from those previously seen.

From the point of view of the seed patterns, we can see in Figure~\ref{fig:distanceEstimatorCorrelation} that, for single seeds, selection done for both {\em values} gives the same optimal seed pattern {\tt 11*1} (or its mirror) for weight 3, and the same optimal seed pattern {\tt 1*1**11} (or its mirror) for weight 4. The choice for the optimal double seed patterns differs between the {\em multi-hit} or {\em coverage values}, and this difference is even more marked for seeds of weight 4.

However, computing the coverage is more difficult than simply counting common $k$-mers. We justify in the next part that, given two easily measurable assumptions on the sequences and the $k$-mer weight, this task can be done efficiently.

\subsubsection{Coverage algorithmic point of view}
\label{subsubsection:CoverageAlgorithm}
In this part, we briefly describe how coverage can be computed efficiently. Given two sequences $s_1$ and $s_2$ of equivalent length, we want to search for the {\em spaced} $k$-mers that are common to $s_1$ and $s_2$. But, more than establishing a frequency profile for these common $k$-mer {\em codes}, the main idea is here to find a set of common $k$-mers that have the same order of position occurrences on $s_1$ and $s_2$. To do so, one solution is to keep occurrences of any of the $4^k$ possible $k$-mers in a {\em reverse list of positions} (given one $k$-mer code, we have two lists of positions where this $k$-mer occurs, on $s_1$ or respectively on $s_2$). Keeping the common $k$-mers of both $s_1$ and $s_2$, sorting their list of pairs of occurrence positions according to the positions of one of the two sequences (for example positions along $s_1$), then applying a LIS (or a windowed LIS if the two sequences are not of similar lengths) on $s_2$, provided that spurious $k$-mers (those occurring randomly) are not frequent, will give a better approximation for the number of true hits, and thus can be used to compute the coverage.

Note first that the LIS can be computed in $t \cdot \log(t)$ time~\citep{SchenstedCJM61} where $t$ is the number of {\em hits} (e.g. pairs of positions for a common $k$-mer) : this value $t$, provided that $k$ is well chosen to correctly filter {\em spurious $k$-mers} and there is no composition bias on both sequences, must be either close to $|s_1|$ and $|s_2|$ if the $s_1$ and $s_2$ sequences are similar (and without self-repetitive bias/low complexity regions), or reasonably low if the sequences are non-similar, but can be otherwise high for low complexity/self-repeating/redundant regions that similarity search tools usually want to avoid.

\begin{figure}[htb]
 \centering
 \caption[Mealy coverage increment automaton]{\label{fig:reverseHitAutomaton}Mealy coverage increment automaton for hits of the seeds $\{\pi_1,\pi_2\} = \{ {\tt 1 1 * 1} , {\tt 1 * 1 * 1} \}$}
 \includegraphics[width=\textwidth]{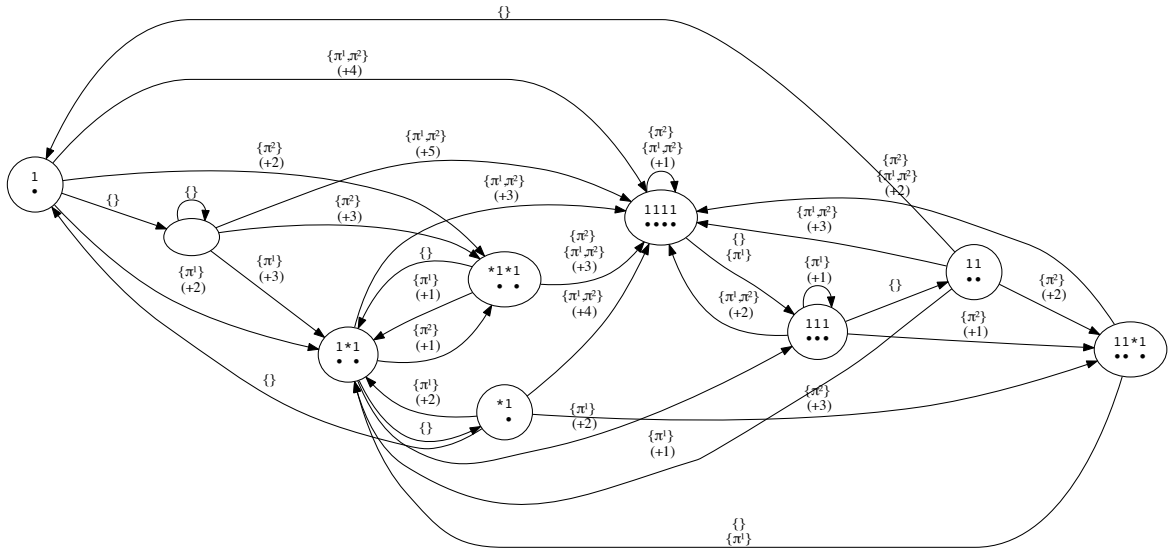}
\end{figure}

Note also that, once the common and ordered hits are collected by the LIS procedure, it is possible to compute the coverage using :
\begin{itemize}
 \item either a masking process using shift-or for collecting the coverage symbols, and then computing the coverage increment (which implies an additional CPU cost if no {\em population count instruction} is available), 
 \item or an automaton (an example is provided in Figure~\ref{fig:reverseHitAutomaton} for the hits of the seeds $\{\pi_1,\pi_2\} = \{ {\tt 1 1 * 1} , {\tt 1 * 1 * 1} \}$) that keeps the last overlapping suffix of the previously encountered hits for any of the seeds. This automaton has an alphabet of size $2^{\# \mathrm{seeds}}$ since we record whether or not there is a seed hit for each seed. Otherwise, a very similar definition to the coverage automaton holds. Once this automaton is built, it is possible to compute the coverage increment in constant time.
\end{itemize}

In both cases, gaps ({\em indels}) must be taken into consideration because they break, from a {\em dot-plot} point of view, {\em diagonals}, thus reinitializing the automaton or the coverage mask.

\section{Concluding Remarks}
\label{section:concludingRemarks}

We have presented how the {\em coverage criterion}~\citep{BensonMakSPIRE08,MartinJSM13} can help in measuring the seed efficiency in two recent problems : a classifier based on spaced $k$-mers~\citep{OnoderaShibuyaMLDM13}, and a $k$-mer alignment-free distance estimation~\citep{BodenEtAlGCB13,LeimeisterEtAlBioinformatics14,HorwegeEtAlNAR14}. We have also shown how to extend the second one to be even more sensitive.

The Moore (or Mealy) automaton obtained to measure the coverage criterion is by itself of interest for several reasons : its size seems to be bounded by $polynom(w,r) \times 3^r$ even if the bound obtained now is rather limited and exponential (see the Appendix).

For example the {\em coverage automaton} size for the PatternHunter 1 seed {\small\tt 111*1**1*1**11*111} is :
\begin{center}
  \begin{tabular}{|C{3cm}|C{3cm}|C{3cm}|C{3cm}|}
    \hline
    \multicolumn{2}{|c|}{Moore} & \multicolumn{2}{c|}{Mealy}\\[-3mm]
    \multicolumn{2}{|c|}{\small{\tt iedera} development version 1.06 $\alpha7$} & \multicolumn{2}{c|}{\small Matlab code + gap-system FR}\\[-1mm]
    \hline
    current size & minimized & current size & minimized \\[-3mm]
    4312 states & 4260 states & 4215 states & 3782 states \\[0mm]
    \hline
  \end{tabular}
\end{center}
\noindent where the {\em current sizes} for Moore and Mealy automata are respectively obtained by the Iedera tool~\citep[][version 1.06 $\alpha7$]{IederaSoftwareUnpublished}, or by the Matlab code~\citep{MartinNoeUn14} before minimization by the gap-system FR package~\citep{GapFR12}. These sizes can be compared with those of the mere {\em multi-hit automaton} :
\begin{center}
  \begin{tabular}{|C{3cm}|C{3cm}|C{3cm}|C{3cm}|}
    \hline
    \multicolumn{2}{|c|}{Moore} & \multicolumn{2}{c|}{Mealy}\\[-3mm]
    \multicolumn{2}{|c|}{\small{\tt iedera} development version 1.06 $\alpha7$} & \multicolumn{2}{c|}{\small Matlab code + gap-system FR}\\[-1mm]
    \hline
    current size & minimized & current size & minimized \\[-3mm]
    322 states & 322 states & 281 states & 278 states \\[0mm]
    \hline
  \end{tabular}
\end{center}
Although the coverage automaton is more than ten times larger than the equivalent multi-hit automaton, it is still usable for dynamic programming computation.

This is even true for {\bf multiple spaced seeds}. For example the {\em coverage automaton} size for the PatternHunter 2 seeds of weight 11 : 
{\small \tt 111*1**1*1**11*111, 1111**11**1*1****1*11, 11*1****11***1*1*1111, 111*111*1***1111} (called {\em first four} in~\cite{PatternHunter04}) is : 
\begin{center}
  \begin{tabular}{|C{3cm}|C{3cm}|C{3cm}|C{3cm}|}
    \hline
    \multicolumn{2}{|c|}{Moore} & \multicolumn{2}{c|}{Mealy}\\[-3mm]
    \multicolumn{2}{|c|}{\small{\tt iedera} development version 1.06 $\alpha7$} & \multicolumn{2}{c|}{\small Matlab code + gap-system FR}\\[-1mm]
    \hline
    current size & minimized & current size & minimized \\[-3mm]
    154412 states & 143736 states & {\em\small not available} & 127049 states \\[0mm]
    \hline
  \end{tabular}
\end{center}
to be compared again with the mere {\em multi-hit automaton} current size (and its minimal size) :
\begin{center}
  \begin{tabular}{|C{3cm}|C{3cm}|C{3cm}|C{3cm}|}
    \hline
    \multicolumn{2}{|c|}{Moore} & \multicolumn{2}{c|}{Mealy}\\[-3mm]
    \multicolumn{2}{|c|}{\small{\tt iedera} development version 1.06 $\alpha7$} & \multicolumn{2}{c|}{\small Matlab code + gap-system FR}\\[-1mm]
    \hline
    current size & minimized & current size & minimized \\[-3mm]
    5119 states & 4963 states & {\em\small not available} & 4183 states \\[0mm]
    \hline
  \end{tabular}
\end{center}
Although more than 20 times larger than the equivalent multi-hit automaton, the coverage automaton for multiple seeds is again still usable for dynamic programming computation. 

It would be also interesting (but out of the scope of this article) to consider SVM kernels or $k$-mer distances with {\em subset seed}~\citep{KucherovNoeRoytbergCIAA07,YangZhangJCB08,GambinEtAlBIOSTEC11,FrithNoeNAR14} or more general {\em vector seed}~\citep{BrejovaBrownVinarJCSS05} techniques. Several string kernels, such as the {\em mismatch string kernel}~\citep{LeslieEtAlBioinformatics04}, use this general concept, but generate full neighborhoods (all the words at a given distance from a given $k$-mer). 
Moreover {\em optimal resolution} [best seed weight]~\citep{SimsaEtALPNAS09} remains an open problem for spaced seeds in both SVM kernels or $k$-mer distance problems. Note also, if one wants to avoid this {\em optimal resolution} question, {\em seed design} and {\em increasing weight} can be combined~\citep[as done in][]{CsurosCPM04,KielbasaWanSatoHortonFrithGENOMERESEARCH11}, but may not be always directly compatible with the aforementioned cited works on variable $k$-mers.

A last idea to explore is also to merge the definition of clumps~\citep{StefanovRobinSchbathDAM07,BassinoClementFayolleNicodemeDMTCS08,MartinColemanJAP11,MarshallHermsKaltenbachRahmannTCBB12,RegnierFangIakovishinaANALCO14} with coverage, for example by giving {\em more significance} (than a linear weight function) to coverage provided by clumps of hits than coverage provided by isolated hits.

\section*{Acknowledgements}
D.E.K. Martin was supported in this research by the National Science Foundation under Grant DMS-1107084.
L. No{\'e}  was supported by a CNRS Mastodons grant, and benefited from a half-time course buyout from the French Institute for Research in Computer Science and Automation (Inria).

\section*{Author Disclosure Statement}
No competing financial interests exist.

{\small
  \bibliographystyle{plainnat}
  \bibliography{coverage-sensitivity}

\begin{thebibliography}{79}
\providecommand{\natexlab}[1]{#1}
\providecommand{\url}[1]{\texttt{#1}}
\expandafter\ifx\csname urlstyle\endcsname\relax
  \providecommand{\doi}[1]{doi: #1}\else
  \providecommand{\doi}{doi: \begingroup \urlstyle{rm}\Url}\fi

\bibitem[Aho and Corasick(1975)]{AhoCorasick74}
Alfred~V. Aho and Margaret~J. Corasick.
\newblock Efficient string matching: An aid to bibliographic search.
\newblock \emph{Communications of the ACM}, 18\penalty0 (6):\penalty0 333--340,
  1975.
\newblock \doi{10.1145/360825.360855}.

\bibitem[Apostolico et~al.(2014)Apostolico, Guerra, and
  Pizzi]{ApostolicoEtAlDCC14}
Alberto Apostolico, Concettina Guerra, and Cinzia Pizzi.
\newblock Alignment free sequence similarity with bounded {H}amming distance.
\newblock In \emph{{P}roceedings of the {D}ata {C}ompression {C}onference
  ({DCC})}, 2014.
\newblock \doi{10.1109/DCC.2014.57}.

\bibitem[Bartholdi(2012)]{GapFR12}
Laurent Bartholdi.
\newblock Functionally recursive groups.
\newblock \url{http://www.gap-system.org/Manuals/pkg/fr-2.1.1/doc/chap0.html},
  2012.

\bibitem[Bassino et~al.(2008)Bassino, Cl{\'e}ment, Fayolle, and
  Nicod{\`e}me]{BassinoClementFayolleNicodemeDMTCS08}
Fr{\'e}d{\'e}rique Bassino, Julien Cl{\'e}ment, Julien Fayolle, and Pierre
  Nicod{\`e}me.
\newblock Constructions for clumps statistics.
\newblock \emph{Discrete Mathematics and Theoretical Computer Science},
  AI:\penalty0 179--194, 2008.

\bibitem[Battaglia et~al.(2009)Battaglia, Cangelosi, Grossi, and
  Pisanti]{BattagliaEtAlTCS09}
Giovanni Battaglia, Davide Cangelosi, Roberto Grossi, and Nadia Pisanti.
\newblock Masking patterns in sequences: A new class of motif discovery with
  don't cares.
\newblock \emph{Theoretical Computer Science}, 410\penalty0 (43):\penalty0
  4327--4340, 2009.
\newblock \doi{10.1016/j.tcs.2009.07.014}.

\bibitem[Benson and Mak(2008)]{BensonMakSPIRE08}
Gary Benson and Denise~Y.F. Mak.
\newblock Exact distribution of a spaced seed statistic for {DNA} homology
  detection.
\newblock In \emph{Proceedings of the {I}nternational {S}ymposium on {S}tring
  {P}rocessing and {I}nformation {R}etrieval ({SPIRE})}, volume 5280 of
  \emph{LNCS}, pages 282--293, 2008.
\newblock \doi{10.1007/978-3-540-89097-3_27}.

\bibitem[Boden et~al.(2013)Boden, Sch\"{o}neich, Horwege, Lindner, Leimeister,
  and Morgenstern]{BodenEtAlGCB13}
Marcus Boden, Martin Sch\"{o}neich, Sebastian Horwege, Sebastian Lindner, Chris
  Leimeister, and Burkhard Morgenstern.
\newblock Alignment-free sequence comparison with spaced $k$-mers.
\newblock In \emph{{P}roceedings of the {G}erman {C}onference on
  {B}ioinformatics ({GCB})}, volume~34 of \emph{OpenAccess Series in
  Informatics (OASIcs)}, pages 24--34, 2013.
\newblock \doi{10.4230/OASIcs.GCB.2013.24}.

\bibitem[Brejov{\'a} et~al.(2005)Brejov{\'a}, Brown, and
  Vina{\v{r}}]{BrejovaBrownVinarJCSS05}
Bro{\v{n}}a Brejov{\'a}, Daniel~G. Brown, and Tom{\'a}{\v{s}} Vina{\v{r}}.
\newblock Vector seeds: An extension to spaced seeds.
\newblock \emph{Journal of Computer and System Sciences}, 70\penalty0
  (3):\penalty0 364--380, 2005.
\newblock \doi{10.1016/j.jcss.2004.12.008}.

\bibitem[B{\v{r}}inda(2014)]{BrindaAFL14}
Karel B{\v{r}}inda.
\newblock Languages of lossless seeds.
\newblock In \emph{Proceedings of the International Conference on {A}utomata
  and {F}ormal {L}anguages ({AFL})}, volume 151, pages 139--150, 2014.
\newblock \doi{10.4204/EPTCS.151.9}.

\bibitem[Buhler et~al.(2005)Buhler, Keich, and Sun]{BuhlerKeichSunJCSS05}
Jeremy Buhler, Uri Keich, and Yanni Sun.
\newblock Designing seeds for similarity search in genomic {DNA}.
\newblock \emph{Journal of Computer and System Sciences}, 70\penalty0
  (3):\penalty0 342--363, 2005.
\newblock \doi{10.1016/j.jcss.2004.12.003}.

\bibitem[Burge et~al.(2012)Burge, Daub, Eberhardt, Tate, Barquist, Nawrocki,
  Eddy, Gardner, and Bateman]{RFAM11}
Sarah~W. Burge, Jennifer Daub, Ruth Eberhardt, John Tate, Lars Barquist,
  Eric~P. Nawrocki, Sean~R. Eddy, Paul~P. Gardner, and Alex Bateman.
\newblock {R}fam 11.0: 10 years of {RNA} families.
\newblock \emph{Nucleic Acids Research}, 41\penalty0 (D1):\penalty0 D226--D232,
  2012.
\newblock \doi{10.1093/nar/gks1005}.

\bibitem[Burkhardt and K{\"a}rkk{\"a}inen(2002)]{BurkhardtKarkkainenFI03}
Stefan Burkhardt and Juha K{\"a}rkk{\"a}inen.
\newblock Better filtering with gapped $q$-grams.
\newblock \emph{Fundamenta Informaticae}, 56\penalty0 (1-2):\penalty0 51--70,
  2002.

\bibitem[Burkhardt et~al.(1999)Burkhardt, Crauser, Ferragina, Lenhof, Rivals,
  and Vingron]{QuasarRECOMB99}
Stefan Burkhardt, Andreas Crauser, Paolo Ferragina, Hans-Peter Lenhof, Eric
  Rivals, and Martin Vingron.
\newblock $q$-gram based database searching using a suffix array {(QUASAR)}.
\newblock In \emph{Proceedings of the Annual International Conference on
  Research in Computational Molecular Biology (RECOMB)}, pages 77--83, 1999.
\newblock \doi{10.1145/299432.299460}.

\bibitem[Chen et~al.(2009)Chen, Souaiaia, and
  Chen]{ChenSouaiaiaChenBioinformatics09}
Yangho Chen, Tate Souaiaia, and Ting Chen.
\newblock {P}er{M}: efficient mapping of short sequencing reads with periodic
  full sensitive spaced seeds.
\newblock \emph{Bioinformatics}, 25\penalty0 (19):\penalty0 2514--2521, 2009.
\newblock \doi{10.1093/bioinformatics/btp486}.

\bibitem[Chor et~al.(2009)Chor, Horn, Goldman, Levy, and
  Massingham]{ChorEtAlGenomeBiology09}
Benny Chor, David Horn, Nick Goldman, Yaron Levy, and Tim Massingham.
\newblock Genomic {DNA} k-mer spectra: Models and modalities.
\newblock \emph{Genome Biology}, 10:\penalty0 R108, 2009.
\newblock \doi{10.1186/gb-2009-10-10-r108}.

\bibitem[Comin and Verzotto(2012)]{CominVerzottoAMB12}
Matteo Comin and Davide Verzotto.
\newblock Alignment-free phylogeny of whole genomes using underlying subwords.
\newblock \emph{Algorithms for Molecular Biology}, 7\penalty0 (34), 2012.
\newblock \doi{10.1186/1748-7188-7-34}.

\bibitem[Cs{\H{u}}r{\"o}s(2004)]{CsurosCPM04}
Mikl{\'o}s Cs{\H{u}}r{\"o}s.
\newblock Performing local similarity searches with variable length seeds.
\newblock In \emph{Proceedings of the 15th Annual Combinatorial Pattern
  Matching Symposium (CPM)}, volume 3109 of \emph{LNCS}, pages 373--387, 2004.
\newblock \doi{10.1007/978-3-540-27801-6_28}.

\bibitem[David et~al.(2011)David, Dzamba, Lister, Ilie, and
  Brudno]{SHRiMP2Bioinformatics11}
Matei David, Misko Dzamba, Dan Lister, Lucian Ilie, and Michael Brudno.
\newblock {SHRiMP2}: Sensitive yet practical short read mapping.
\newblock \emph{Bioinformatics}, 27\penalty0 (7):\penalty0 1011--1012, 2011.
\newblock \doi{10.1093/bioinformatics/btr046}.

\bibitem[Didier et~al.(2012)Didier, Corel, Laprevotte, Grossmann, and
  Land{\`e}s-Devauchelle]{DidierEtAlTCS12}
Gilles Didier, Eduardo Corel, Ivan Laprevotte, Alex Grossmann, and Claudine
  Land{\`e}s-Devauchelle.
\newblock Variable length local decoding and alignment-free sequence
  comparison.
\newblock \emph{Theoretical Computer Science}, 462:\penalty0 1--11, 2012.
\newblock \doi{10.1016/j.tcs.2012.08.005}.

\bibitem[Do~Duc et~al.(2012)Do~Duc, Dinh, Dang, Laukens, and
  Hoang]{AcoSeedANTS12}
Dong Do~Duc, Huy~Q. Dinh, Thanh~Hai Dang, Kris Laukens, and Xuan~Huan Hoang.
\newblock {A}co{S}ee{D}: An ant colony optimization for finding optimal spaced
  seeds in biological sequence search.
\newblock In \emph{Proceedings of the 8th International Conference on Swarm
  Intelligence (ANTS)}, volume 7461 of \emph{LNCS}, pages 204--211, 2012.
\newblock \doi{10.1007/978-3-642-32650-9_19}.

\bibitem[Edgar(2004)]{Muscle04}
Robert~C. Edgar.
\newblock {MUSCLE}: Multiple sequence alignment with high accuracy and high
  throughput.
\newblock \emph{Nucleic Acids Research}, 32\penalty0 (5):\penalty0 1792--1797,
  2004.
\newblock \doi{10.1093/nar/gkh340}.

\bibitem[Egidi and Manzini(2014{\natexlab{a}})]{EgidiManziniFI14}
Lavinia Egidi and Giovanni Manzini.
\newblock Spaced seeds design using perfect rulers.
\newblock \emph{Fundamenta Informaticae}, 131\penalty0 (2):\penalty0 187--203,
  2014{\natexlab{a}}.
\newblock \doi{10.3233/FI-2014-1009}.

\bibitem[Egidi and Manzini(2014{\natexlab{b}})]{EgidiManziniTCS14}
Lavinia Egidi and Giovanni Manzini.
\newblock Design and analysis of periodic multiple seeds.
\newblock \emph{Theoretical Computer Science}, 522:\penalty0 62--76,
  2014{\natexlab{b}}.
\newblock \doi{10.1016/j.tcs.2013.12.007}.

\bibitem[Farach-Colton et~al.(2007)Farach-Colton, Landau, Cenk~Sahinalp, and
  Tsur]{FarachEtAlJCSS07}
Martin Farach-Colton, Gad~M. Landau, S{\"u}leyman Cenk~Sahinalp, and Dekel
  Tsur.
\newblock Optimal spaced seeds for faster approximate string matching.
\newblock \emph{Journal of Computer and System Sciences}, 73\penalty0
  (7):\penalty0 1035--1044, 2007.
\newblock \doi{10.1016/j.jcss.2007.03.007}.

\bibitem[Frith and No{\'e}(2014)]{FrithNoeNAR14}
Martin~C. Frith and Laurent No{\'e}.
\newblock Improved search heuristics find 20 000 new alignments between human
  and mouse genomes.
\newblock \emph{Nucleic Acids Research}, 42\penalty0 (7):\penalty0 e59, 2014.
\newblock \doi{10.1093/nar/gku104}.

\bibitem[Gambin et~al.(2011)Gambin, Lasota, Startek, Sykulski, No{\'e}, and
  Kucherov]{GambinEtAlBIOSTEC11}
Anna Gambin, S{\l}awomir Lasota, Micha{\l} Startek, Macieij Sykulski, Laurent
  No{\'e}, and Gregory Kucherov.
\newblock Subset seed extension to {P}rotein {BLAST}.
\newblock In \emph{Proceedings of the International Conference on
  Bioinformatics Models, Methods and Algorithms}, pages 149--158.
  {S}ci{T}e{P}ress Digital Library, 2011.
\newblock \doi{10.5220/0003147601490158}.

\bibitem[Ghandi et~al.(2014{\natexlab{a}})Ghandi, Lee, Mohammad-Noori, and
  Beer]{GhandiEtAlPLoSComputationalBiology14}
Mahmoud Ghandi, Dongwon Lee, Morteza Mohammad-Noori, and Michael~A. Beer.
\newblock Enhanced regulatory sequence prediction using gapped k-mer features.
\newblock \emph{{PLoS} Computational Biology}, 10\penalty0 (7):\penalty0
  e1003711, July 2014{\natexlab{a}}.
\newblock \doi{10.1371/journal.pcbi.1003711}.

\bibitem[Ghandi et~al.(2014{\natexlab{b}})Ghandi, Mohammad-Noori, and
  Beer]{GhandiEtAlJMB14}
Mahmoud Ghandi, Morteza Mohammad-Noori, and Michael~A. Beer.
\newblock Robust k-mer frequency estimation using gapped k-mers.
\newblock \emph{Journal of Mathematical Biology}, 69\penalty0 (2):\penalty0
  469--500, August 2014{\natexlab{b}}.
\newblock \doi{10.1007/s00285-013-0705-3}.

\bibitem[Giladi et~al.(2010)Giladi, Healy, Myers, Hart, Kapranov, Lipson,
  Roels, Thayer, and Letovsky]{GiladiEtAlJCB10}
Eldar Giladi, John Healy, Gene Myers, Chris Hart, Phillip Kapranov, Doron
  Lipson, Steven Roels, Edward Thayer, and Stan Letovsky.
\newblock Error tolerant indexing and alignment of short reads with covering
  template families.
\newblock \emph{Journal of Computational Biology}, 17\penalty0 (10):\penalty0
  1397--1411, 2010.
\newblock \doi{10.1089/cmb.2010.0005}.

\bibitem[Giraud et~al.(2014)Giraud, Salson, Duez, Villenet, Quief, Caillault,
  Grardel, Roumier, Preudhomme, and Figeac]{VidjilBMCGenomics14}
Mathieu Giraud, Mika\"el Salson, Marc Duez, C\'eline Villenet, Sabine Quief,
  Aur\'elie Caillault, Nathalie Grardel, Christophe Roumier, Claude Preudhomme,
  and Martin Figeac.
\newblock Fast multiclonal clusterization of {V(D)J} recombinations from
  high-throughput sequencing.
\newblock \emph{{BMC} Genomics}, 15\penalty0 (409), 2014.
\newblock \doi{10.1186/1471-2164-15-409}.

\bibitem[Harris(2007)]{HarrisLASTZPhD07}
Robert~S. Harris.
\newblock \emph{Improved pairwise alignment of genomic DNA}.
\newblock Ph.d. thesis, The Pennsylvania State University, December 2007.

\bibitem[Haubold et~al.(2005)Haubold, Pierstorff, M\"oller, and
  Wiehe]{HauboldEtAlBMCBioinformatics05}
Bernhard Haubold, Nora Pierstorff, Friedrich M\"oller, and Thomas Wiehe.
\newblock Genome comparison without alignment using shortest unique substrings.
\newblock \emph{{BMC} {B}ioinformatics}, 6\penalty0 (123), 2005.
\newblock \doi{10.1186/1471-2105-6-123}.

\bibitem[Homer et~al.(2009)Homer, Merriman, and Nelson]{BFAST09}
Nils Homer, Barry Merriman, and Stanley~F. Nelson.
\newblock {BFAST}: An alignment tool for large scale genome resequencing.
\newblock \emph{{PLoS One}}, 4\penalty0 (11):\penalty0 e7767, 2009.
\newblock \doi{10.1371/ journal.pone.0007767}.

\bibitem[Hopcroft(1971)]{Hopcroft71}
John Hopcroft.
\newblock An $n \log n$ algorithm for minimizing the states in a finite
  automaton.
\newblock In Z.~Kohavi and A.~Paz, editors, \emph{The Theory of Machines and
  Computation}, pages 189--196. Academic Press, New York, 1971.

\bibitem[Horwege et~al.(2014)Horwege, Lindner, Boden, Hatje, Kollmar,
  Leimeister, and Morgenstern]{HorwegeEtAlNAR14}
Sebastian Horwege, Sebastian Lindner, Marcus Boden, Klas Hatje, Martin Kollmar,
  Chris-Andr{\'e} Leimeister, and Burkhard Morgenstern.
\newblock Spaced words and kmacs: Fast alignment-free sequence comparison based
  on inexact word matches.
\newblock \emph{Nucleic Acids Research}, 42\penalty0 (W1):\penalty0 W7--W11,
  2014.
\newblock \doi{10.1093/nar/gku398}.

\bibitem[Huang(2006)]{HuangDPRing06}
Liang Huang.
\newblock Dynamic programming algorithms in semiring and hypergraph frameworks.
\newblock Technical report, University of Pennsylvania, Philadelphia, USA,
  November 2006.

\bibitem[Ilie et~al.(2011)Ilie, Ilie, and
  Mansouri~Bigvand]{IlieEtAlBioinformatics11}
Lucian Ilie, Silvana Ilie, and Anahita Mansouri~Bigvand.
\newblock {SpEED}: fast computation of sensitive spaced seeds.
\newblock \emph{Bioinformatics}, 27\penalty0 (17):\penalty0 2433--2434, 2011.
\newblock \doi{10.1093/bioinformatics/btr368}.

\bibitem[Ilie et~al.(2013)Ilie, Mohamadi, Brian~Golding, and
  Smyth]{BONDBMCBioinformatics13}
Lucian Ilie, Hamid Mohamadi, Geoffrey Brian~Golding, and William~F. Smyth.
\newblock {BOND}: {B}asic {O}ligo{N}ucleotide {D}esign.
\newblock \emph{{BMC} {B}ioinformatics}, 14\penalty0 (69), 2013.
\newblock \doi{10.1186/1471-2105-14-69}.

\bibitem[Joachims(2002)]{Thorsten02}
Thorsten Joachims.
\newblock \emph{{L}earning to {C}lassify {T}ext using {S}upport {V}ector
  {M}achines}.
\newblock Kluwer/Springer, 2002.
\newblock \doi{10.1007/978-1-4615-0907-3}.

\bibitem[Keich et~al.(2004)Keich, Li, Ma, and Tromp]{KeichLiMaTrompDAM04}
Uri Keich, Ming Li, Bin Ma, and John Tromp.
\newblock On spaced seeds for similarity search.
\newblock \emph{Discrete Applied Mathematics}, 138\penalty0 (3):\penalty0
  253--263, 2004.
\newblock \doi{10.1016/S0166-218X(03)00382-2}.

\bibitem[Kie{\l}basa et~al.(2011)Kie{\l}basa, Wan, Sato, Horton, and
  Frith]{KielbasaWanSatoHortonFrithGENOMERESEARCH11}
Szymon~M. Kie{\l}basa, Raymond Wan, Kengo Sato, Paul Horton, and Martin~C.
  Frith.
\newblock Adaptive seeds tame genomic sequence comparison.
\newblock \emph{Genome Research}, 21\penalty0 (3):\penalty0 487--493, 2011.
\newblock \doi{10.1101/gr.113985.110}.

\bibitem[Kuang et~al.(2005)Kuang, Ie, Wang, Wang, Siddiqi, Freund, and
  Leslie]{KuangEtAlJBCB05}
Rui Kuang, Eugene Ie, Ke~Wang, Kai Wang, Mahira Siddiqi, Yoav Freund, and
  Christina Leslie.
\newblock Profile-based string kernels for remote homology detection and motif
  extraction.
\newblock \emph{Journal of Bioinformatics and Computational Biology},
  3\penalty0 (3):\penalty0 527--550, 2005.
\newblock \doi{10.1109/CSB.2004.135}.

\bibitem[Kucherov et~al.(2005)Kucherov, No{\'e}, and
  Roytberg]{KucherovNoeRoytbergTCBB05}
Gregory Kucherov, Laurent No{\'e}, and Mikhail~A. Roytberg.
\newblock Multiseed lossless filtration.
\newblock \emph{IEEE/ACM Transactions on Computational Biology and
  Bioinformatics (TCBB)}, 2\penalty0 (1):\penalty0 51--61, 2005.
\newblock \doi{10.1109/tcbb.2005.12}.

\bibitem[Kucherov et~al.(2006)Kucherov, No{\'e}, and
  Roytberg]{KucherovNoeRoytbergJBCB06}
Gregory Kucherov, Laurent No{\'e}, and Mikhail~A. Roytberg.
\newblock A unifying framework for seed sensitivity and its application to
  subset seeds.
\newblock \emph{Journal of Bioinformatics and Computational Biology},
  4\penalty0 (2):\penalty0 553--569, 2006.
\newblock \doi{10.1142/S0219720006001977}.

\bibitem[Kucherov et~al.(2007)Kucherov, No{\'e}, and
  Roytberg]{KucherovNoeRoytbergCIAA07}
Gregory Kucherov, Laurent No{\'e}, and Mikhail~A. Roytberg.
\newblock Subset seed automaton.
\newblock In \emph{Proceedings of the 12th International {C}onference on
  {I}mplementation and {A}pplication of {A}utomata ({CIAA})}, volume 4783 of
  \emph{LNCS}, pages 180--191, 2007.
\newblock \doi{10.1007/978-3-540-76336-9_18}.

\bibitem[Kucherov et~al.(2014)Kucherov, No{\'e}, and
  Roytberg]{IederaSoftwareUnpublished}
Gregory Kucherov, Laurent No{\'e}, and Mikhail~A. Roytberg.
\newblock Iedera subset seed design tool.
\newblock \url{http://bioinfo.lifl.fr/yass/iedera.php}, 2014.

\bibitem[Leimeister et~al.(2014)Leimeister, Boden, Horwege, Lindner, and
  Morgenstern]{LeimeisterEtAlBioinformatics14}
Chris-Andr{\'e} Leimeister, Marcus Boden, Sebastian Horwege, Sebastian Lindner,
  and Burkhard Morgenstern.
\newblock Fast alignment-free sequence comparison using spaced-word
  frequencies.
\newblock \emph{Bioinformatics}, 30\penalty0 (14):\penalty0 1991--1999, 2014.
\newblock \doi{10.1093/bioinformatics/btu177}.

\bibitem[Leslie et~al.(2002)Leslie, Eskin, and Stafford~Noble]{LeslieEtAlPSB02}
Christina~S. Leslie, Eleazar Eskin, and William Stafford~Noble.
\newblock The spectrum kernel: A string kernel for {SVM} protein
  classification.
\newblock In \emph{{P}roceedings of the {P}acific {S}ymposium on {B}iocomputing
  ({PSB})}, pages 564--575, 2002.

\bibitem[Leslie et~al.(2004)Leslie, Eskin, Cohen, Weston, and
  Stafford~Noble]{LeslieEtAlBioinformatics04}
Christina~S. Leslie, Eleazar Eskin, Adiel Cohen, Jason Weston, and William
  Stafford~Noble.
\newblock Mismatch string kernels for discriminative protein classification.
\newblock \emph{Bioinfomatics}, 20\penalty0 (4):\penalty0 467--476, 2004.
\newblock \doi{10.1093/bioinformatics/btg431}.

\bibitem[Li et~al.(2004)Li, Ma, Kisman, and Tromp]{PatternHunter04}
Ming Li, Bin Ma, Derek Kisman, and John Tromp.
\newblock {P}attern{H}unter {II}: Highly sensitive and fast homology search.
\newblock \emph{Journal of Bioinformatics and Computational Biology},
  2\penalty0 (3):\penalty0 417--439, 2004.
\newblock \doi{10.1142/S0219720004000661}.

\bibitem[Lin et~al.(2008)Lin, Zhang, Zhang, Ma, and
  Li]{LinZhangZhangMaLiBioinformatics08}
Hao Lin, Zefeng Zhang, Michael~Q. Zhang, Bin Ma, and Ming Li.
\newblock {ZOOM!} {Z}illions {O}f {O}ligos {M}apped.
\newblock \emph{Bioinformatics}, 24\penalty0 (21):\penalty0 2431--2437, 2008.
\newblock \doi{10.1093/bioinformatics/btn416}.

\bibitem[Liu et~al.(2008)Liu, DeSantis, Andersen, and Knight]{LiuEtALNAR08}
Zongzhi Liu, Todd~Z. DeSantis, Gary~L. Andersen, and Rob Knight.
\newblock Accurate taxonomy assignments from 16{S} r{RNA} sequences produced by
  highly parallel pyrosequencers.
\newblock \emph{Nucleic Acids Research}, 36\penalty0 (18):\penalty0 e120, 2008.
\newblock \doi{10.1093/nar/gkn491}.

\bibitem[Lodhi et~al.(2002)Lodhi, Saunders, Shawe-Taylor, Cristianini, and
  Watkins]{LodhiEtAlJMLR02}
Huma Lodhi, Craig Saunders, John Shawe-Taylor, Nello Cristianini, and Chris
  Watkins.
\newblock Text classification using string kernels.
\newblock \emph{Journal of Machine Learning Research}, 2:\penalty0 419--444,
  2002.
\newblock \doi{10.1162/153244302760200687}.

\bibitem[Marschall et~al.(2012)Marschall, Herms, Kaltenbach, and
  Rahmann]{MarshallHermsKaltenbachRahmannTCBB12}
Tobias Marschall, Inke Herms, Hans-Michael Kaltenbach, and Sven Rahmann.
\newblock Probabilistic arithmetic automata and their applications.
\newblock \emph{{IEEE/ACM} Transactions on Computational Biology and
  Bioinformatics ({TCBB})}, 9\penalty0 (6):\penalty0 1737--1750, 2012.
\newblock \doi{10.1109/TCBB.2012.109}.

\bibitem[Martin(2013)]{MartinJSM13}
Donald E.~K. Martin.
\newblock Coverage of spaced seeds as a measure of clumping.
\newblock In \emph{{JSM} {P}roceedings, {S}tatistical {C}omputing Section},
  Alexandria, Virginia, 2013. American Statistical Association.

\bibitem[Martin and Coleman(2011)]{MartinColemanJAP11}
Donald E.~K. Martin and Deidra~A. Coleman.
\newblock Distribution of clump statistics for a collection of words.
\newblock \emph{Journal of Applied Probability}, 48\penalty0 (4):\penalty0
  901--1204, 2011.
\newblock \doi{10.1239/jap/1324046018}.

\bibitem[Martin and No{\'e}(2014)]{MartinNoeUn14}
Donald E.~K. Martin and Laurent No{\'e}.
\newblock Faster exact probabilities for statistics of overlapping pattern
  occurrences.
\newblock \emph{Submitted to the {A}nnals of the {I}nstitute of {S}tatistical
  {M}athematics ({AISM})}, 2014.

\bibitem[Maurer-Stroh et~al.(2013)Maurer-Stroh, Gunalan, Wong, and
  Eisenhaber]{MaurerStrohEtAlJBCB13}
Sebastian Maurer-Stroh, Vithiagaran Gunalan, Wing-Cheong Wong, and Frank
  Eisenhaber.
\newblock A simple shortcut to unsupervised alignment-free phylogenetic genome
  groupings, even from unassembled sequencing reads.
\newblock \emph{Journal of Bioinformatics and Computational Biology},
  11\penalty0 (6):\penalty0 1343005, 2013.
\newblock \doi{10.1142/S0219720013430051}.

\bibitem[Michener and Sokal(1957)]{MichenerSokalEvolution57}
Charles~D. Michener and Robert~R. Sokal.
\newblock A quantitative approach to a problem in classification.
\newblock \emph{Evolution}, 11\penalty0 (2):\penalty0 130--162, June 1957.

\bibitem[Mohri(2009)]{MohriHWAChapter09}
Mehryar Mohri.
\newblock \emph{{H}andbook of {W}eighted {A}utomata}, chapter {W}eighted
  {A}utomata {A}lgorithms, pages 213--254.
\newblock Springer, 2009.
\newblock \doi{10.1007/978-3-642-01492-5_6}.

\bibitem[Nicolas and Rivals(2008)]{NicolasRivalsJCSS08}
Fran{\c{c}}ois Nicolas and {\'E}ric Rivals.
\newblock Hardness of optimal spaced seed design.
\newblock \emph{Journal of Computer and System Sciences}, 74\penalty0
  (5):\penalty0 831--849, 2008.
\newblock \doi{10.1016/j.jcss.2007.10.001}.

\bibitem[Nuel(2008)]{NuelJAP08}
Gr{\'e}gory Nuel.
\newblock Pattern {M}arkov chains: optimal {M}arkov chain embedding through
  deterministic finite automata.
\newblock \emph{Journal of Applied Probability}, 45:\penalty0 226--243, 2008.

\bibitem[Nuel(2011)]{NuelBTMChapter11}
Gr{\'e}gory Nuel.
\newblock \emph{{B}ioinformatics - {T}rends and {M}ethodologies}, chapter
  {S}ignificance {S}core of {M}otifs in {B}iological {S}equences.
\newblock {InTech}, 2011.
\newblock \doi{10.5772/18448}.

\bibitem[{Octave community}(2014)]{Octave14}
{Octave community}.
\newblock {GNU Octave 3.8}.
\newblock \url{http://www.gnu.org/software/octave/}, 2014.

\bibitem[Onodera and Shibuya(2013)]{OnoderaShibuyaMLDM13}
Taku Onodera and Tetsuo Shibuya.
\newblock The gapped spectrum kernel for support vector machines.
\newblock In \emph{{P}roceedings of the {I}nternational {C}onference on
  {M}achine {L}earning and {D}ata {M}ining in {P}attern {R}ecognition (MLDM)},
  volume 7988 of \emph{LNCS}, pages 1--15, 2013.
\newblock \doi{10.1007/978-3-642-39712-7_1}.

\bibitem[Pin(1998)]{Pin98}
Jean-\'Eric Pin.
\newblock Tropical semirings.
\newblock In J.~Gunawardena, editor, \emph{Idempotency}, volume~11 of
  \emph{Publ. Newton Inst.}, pages 50--69, Bristol, 1998. Cambridge Univ.
  Press.

\bibitem[Qi et~al.(2004)Qi, Luo, and Hao]{CVTree04}
Ji~Qi, Hong Luo, and Bailin Hao.
\newblock {CVT}ree: A phylogenetic tree reconstruction tool based on whole
  genomes.
\newblock \emph{Nucleic Acids Research}, 32\penalty0 (suppl 2):\penalty0
  W45--W47, 2004.
\newblock \doi{10.1093/nar/gkh362}.

\bibitem[Rasmussen et~al.(2006)Rasmussen, Stoye, and Myers]{SwiftJCB06}
Kim~R. Rasmussen, Jens Stoye, and Eugene~W. Myers.
\newblock Efficient $q$-gram filters for finding all $\epsilon$-matches over a
  given length.
\newblock \emph{Journal of Computational Biology}, 13\penalty0 (2):\penalty0
  296--308, 2006.
\newblock \doi{10.1089/cmb.2006.13.296}.

\bibitem[R{\'e}gnier et~al.(2014)R{\'e}gnier, Fang, and
  Iakovishina]{RegnierFangIakovishinaANALCO14}
Mireille R{\'e}gnier, Billy Fang, and Daria Iakovishina.
\newblock Clump combinatorics, automata, and word asymptotics.
\newblock In \emph{{P}roceedings of the {W}orkshop on {A}nalytic {A}lgorithmics
  and {C}ombinatorics ({ANALCO})}, 2014.
\newblock \doi{10.1137/1.9781611973204.6}.

\bibitem[Saigo et~al.(2004)Saigo, Vert, Ueda, and
  Akutsu]{SaigoEtAlBioinformatics04}
Hiroto Saigo, Jean-Philippe Vert, Nobuhisa Ueda, and Tatsuya Akutsu.
\newblock Protein homology detection using string alignment kernels.
\newblock \emph{Bioinfomatics}, 20\penalty0 (11):\penalty0 1682--1689, 2004.
\newblock \doi{10.1093/bioinformatics/bth141}.

\bibitem[Schensted(1961)]{SchenstedCJM61}
Craige Schensted.
\newblock Longest increasing and decreasing subsequences.
\newblock \emph{Canadian Journal of Mathematics}, 13:\penalty0 179--191, 1961.
\newblock \doi{10.4153/CJM-1961-015-3}.

\bibitem[Simon(1988)]{Simon88}
Imre Simon.
\newblock Recognizable sets with multiplicities in the tropical semiring.
\newblock In \emph{Mathematical foundations of computer science}, volume 324 of
  \emph{LNCS}, pages 107--120, 1988.
\newblock \doi{10.1007/BFb0017135}.

\bibitem[Simsa et~al.(2009)Simsa, Juna, Wua, and Kim]{SimsaEtALPNAS09}
Gregory~E. Simsa, Se-Ran Juna, Guohong~A. Wua, and Sung-Hou Kim.
\newblock Alignment-free genome comparison with feature frequency profiles
  ({FFP}) and optimal resolutions.
\newblock \emph{{P}roceedings of the {N}ational {A}cademy of {S}ciences},
  106\penalty0 (8):\penalty0 2677--2682, 2009.
\newblock \doi{10.1073/pnas.0813249106}.

\bibitem[Stefanov et~al.(2007)Stefanov, Robin, and
  Schbath]{StefanovRobinSchbathDAM07}
Valeri~T. Stefanov, St{\'e}phane Robin, and Sophie Schbath.
\newblock Waiting times for clumps of patterns and for structured motifs in
  random sequences.
\newblock \emph{Discrete Applied Mathematics}, 155\penalty0 (6-7):\penalty0
  868--880, 2007.
\newblock \doi{10.1016/j.dam.2005.07.016}.

\bibitem[Stropea and Moriyama(2007)]{StropeaMoriyamaGenomics07}
Pooj~K. Stropea and Etsuko~N. Moriyama.
\newblock Simple alignment-free methods for protein classification: A case
  study from {G}-protein-coupled receptors.
\newblock \emph{Genomics}, 89\penalty0 (5):\penalty0 602–--612, 2007.
\newblock \doi{10.1016/j.ygeno.2007.01.008}.

\bibitem[Vinga(2014)]{VingaBriefingsinBioinformatics14}
Susana Vinga.
\newblock Editorial: Alignment-free methods in computational biology.
\newblock \emph{Briefings in Bioinformatics}, 15\penalty0 (3):\penalty0
  341--342, 2014.
\newblock \doi{10.1093/bib/bbu005}.

\bibitem[Vinga and Almeida(2003)]{VingaAlmeidaBioinformatics03}
Susana Vinga and Jonas Almeida.
\newblock Alignment-free sequence comparison - a review.
\newblock \emph{Bioinformatics}, 19\penalty0 (4):\penalty0 513--523, 2003.
\newblock \doi{10.1093/bioinformatics/btg005}.

\bibitem[Yang and Zhang(2008)]{YangZhangJCB08}
Jialiang Yang and Louxin Zhang.
\newblock Run probabilities of seed-like patterns and identifying good
  transition seeds.
\newblock \emph{Journal of Computational Biology}, 15\penalty0 (10):\penalty0
  1295--1313, 2008.
\newblock \doi{10.1089/cmb.2007.0209}.

\bibitem[Zhou et~al.(2010)Zhou, Mihai, and Florea]{ZhouMihaiFloreaCIS10}
Leming Zhou, Ingrid Mihai, and Liliana Florea.
\newblock Spaced seeds for cross-species {cDNA}-to-genome sequence alignment.
\newblock \emph{Communications in Information and Systems}, 10\penalty0
  (2):\penalty0 115--136, 2010.

\end{thebibliography}
}



\newpage
\appendix
\section{Seed coverage automaton size}

We consider in this part the size of the seed automaton. Given a seed of weight $w$ and $r$ jokers we are particularly interested in a bound for the size of the coverage automaton, 
as this can provide a limit on memory needed for future analyses.

In this section, we first solve the problem in the special case of a seed of the form {\tt 1*${}^r$1}, before going to a more general case of a seed of weight $w$ and $r$ jokers, for which we show a more general (but less satisfying) upper bound.

\subsection{Seed {\tt 1*${}^r$1} coverage automaton size}
\label{subsection:CoverageSize101}

\begin{figure}[htb]
 \centering
 \caption[Moore multi-hit and Moore coverage automata sizes]{\label{fig:multihitAndCoverageAutomataSizes}
 Moore multi-hit (a) and Moore coverage (b) automata size illustrated for the seed $\pi = {\tt 1*1}$.\newline
 \medskip
 
{\centering\small\it In boxes are set all the seed prefixes $q$ that can be reached for the Moore multi-hit (a) and the Moore coverage (b) automata. Additionally, on the coverage automaton (b), for each prefix $q$, we have enumerated all the possible coverage strings $c$ that are compatible to form $\langle\substack{q^{} \\ c^{}}\rangle$ states : this is done by substituting any non-covered ${\tt 1}$ symbol of $\langle\substack{q^{} \\ c^{}}\rangle$ (but the last) by a possibly covered one $\underset{\bullet}{\tt 1}$ and, for final states, by considering newly covered positions $\underset{\circ}{\tt 1}$.}
}
 \includegraphics[width=.8\textwidth]{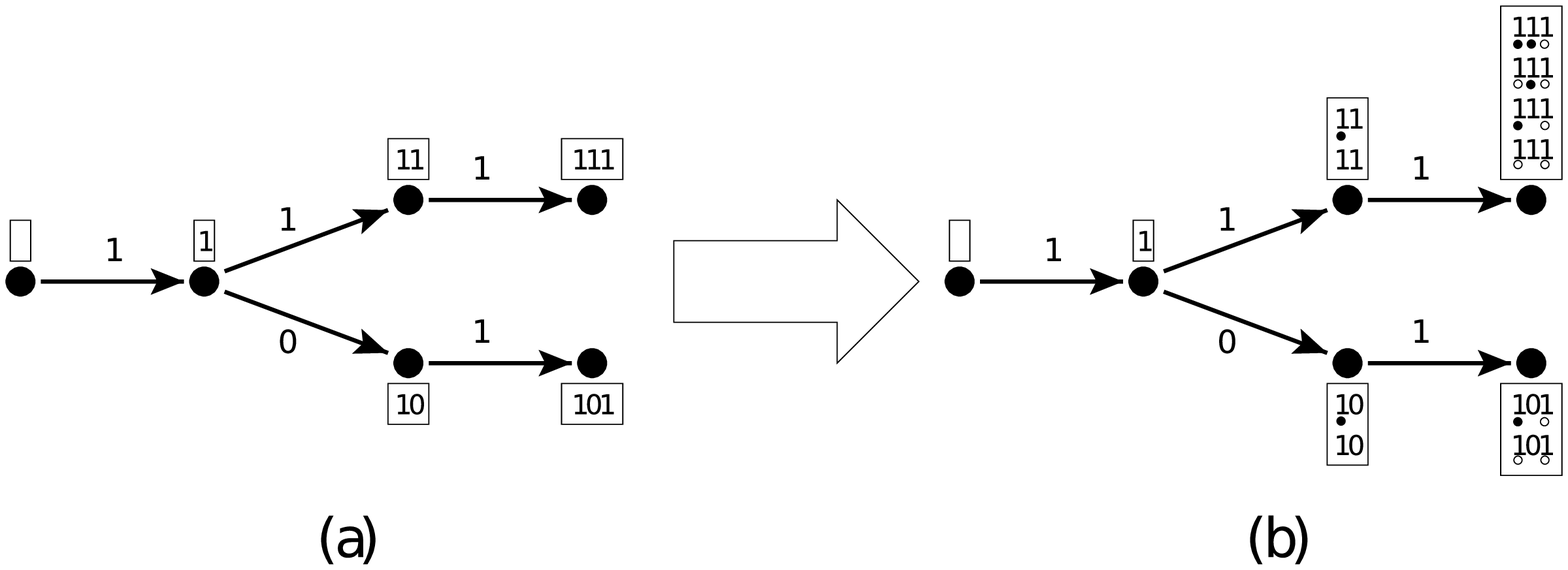}
\end{figure}

The {\tt 1*${}^r$1} seed family has already been shown to reach the multi-hit automaton size bound~\citep{KucherovNoeRoytbergJBCB06} : as a nightmare for the classical seed design tools, such seeds are good candidates to start with.

The {\em multi-hit} automaton size is, in the general case, of maximal size $(w+1)2^{r}$~\citep{BuhlerKeichSunJCSS05,KucherovNoeRoytbergJBCB06}. Moreover, for seeds of the form {\tt 1*${}^r$1}, this size cannot be reduced further~\citep{KucherovNoeRoytbergJBCB06} : thus, {\tt 1*${}^r$1} always have multi-hit automata of size $3 \times 2^r$ (illustrated in Figure~\ref{fig:multihitAndCoverageAutomataSizes} (a) where not all the transitions are shown).

\medskip

The {\em coverage} automaton size for seeds of the form {\tt 1*${}^r$1} is respectively $4 \times 3^r$ for the Moore automaton, and $3 \times 3^r$ for the Mealy automaton.

\medskip

\begin{proof}\itshape
We concentrate first on the Moore automaton. The set of states for the {\em coverage} automaton can be easily deduced from the {\em multi-hit} automaton by considering, for each of the {\em multi-hit} prefixes $q$, all the possible coverages $c$ that are compatible with the current prefix to form reachable $\langle\substack{q^{} \\ c^{}}\rangle$ states. Any prefix $q$ may have any of its {\tt 1}-positions (but the last) covered by a previous hit of a seed if this previous hit {\bf ends} at this {\tt 1}-position (illustrated by the dot symbols of Figure~\ref{fig:multihitAndCoverageAutomataSizes} (b) to mark {\tt 1}-positions already covered). Moreover, it must be noticed that coverage of any {\tt 1}-positions inside $q$ can be chosen independently, by making/disabling a previous hit of a seed using its {\bf first} {\tt 1}-position (this position is not shown on the automaton, thus not overlapping the current prefix, and does not have any side effect). Thus all the possible {\tt 1}-positions (but the last) of a given proper prefix $q$ can be chosen independently with or without coverage.
Thus, for any proper prefix $q$ of length $l+1$ ($0 < l+1 < k $) ($q$ overlaps the first {\em must match} symbol, followed by $l = 1 \ldots r$ {\em joker} symbols of the seed ongoing hit)
\begin{enumerate}
 \item the very first ${\tt 1}$ symbol under a {\em must match} symbol can be covered or not (two possibilities : ${\tt 1}$ or $\underset{\bullet}{\tt 1}$), 
 \item the next $l-1$ symbols under joker symbols can be independently chosen as ${\tt 0}$ or ${\tt 1}$ (three possibilities : ${\tt 0}$, ${\tt 1}$ or $\underset{\bullet}{\tt 1}$), 
 \item the very last symbol under the last joker symbol can be independently chosen as ${\tt 0}$ or ${\tt 1}$ (two possibilities). Note that this ${\tt 1}$, as new, cannot be covered by a previous hit.
\end{enumerate}
For a given prefix $q$ with $l = 1 \ldots r$ jokers, there are thus $2 \times 3^{l-1} \times 2 = 4 \times 3^{l-1}$ possible $\langle\substack{q^{} \\ c^{}}\rangle$ states . Finally, the final states can be seen as prefixes $q$ of length $k=r+2$, where the last $1$ is always newly covered (one choice : $\underset{\circ}{\tt 1}$), the $r$ jokers can be any of ${\tt 0}$, ${\tt 1}$ or $\underset{\bullet}{\tt 1}$ (3 choices), and the very first ${\tt 1}$ can be previously covered or newly covered (2 choices : $\underset{\bullet}{\tt 1}$ or $\underset{\circ}{\tt 1}$ when considering the Moore automaton), leading to $3^r \times 2$ final $\langle\substack{q^{} \\ c^{}}\rangle$ states. At the end, adding the initial state for $q=\epsilon$, and its next state (for $q = ``1"$ corresponding to the first must match position of the seed which cannot be covered), gives :
\[
\underbrace{2}_{\textrm{initial state + next state}}
\quad + \quad
\underbrace{\sum_{l=1}^{r} {4 \times 3^{l-1}}}_{\textrm{proper prefixes of length $0<l+1<k$}}
\quad + \quad
\underbrace{2 \times 3^{r}}_{\textrm{last final states}}
\]
\[=\]
\[
 4 \times 3^{r}
\]

Such seeds {\tt 1*${}^r$1} have thus a Moore coverage automata of size $4 \times 3^r$.

\medskip

 Note that this size cannot be reduced. In other words, given any pair of states $\langle\substack{q_{a} \\ c_{a}}\rangle$ and $\langle\substack{q_{b} \\ c_{b}}\rangle$ on this automaton, and starting (from each of these states) a walk by reading the same (given) string $u$ :
\begin{itemize}
\item if $q_{a}$ and $q_{b}$ are different, then it is always possible to find one string $u$ such that {\em only one of the two} walks reaches a final state~\citep[as done in][]{KucherovNoeRoytbergJBCB06}.
\item otherwise, the coverages $c_{a}$ and $c_{b} $ must be different : it is then always possible to find one string $u$ going to two $final$ states that have a {\em different coverage} increment for the Moore automaton.
\end{itemize}

\medskip

We concentrate now on the Mealy automaton. The main difference with the Moore automaton is that suffixes of full length $k$ (that are final states of the Moore automaton) are not represented because coverage values are set on transitions, and not on states \citep[see][]{MartinNoeUn14}.

For the Mealy automaton of~\cite{MartinNoeUn14}, and seeds of the form {\tt 1*${}^r$1} (of length $k = r + 2$ and weight $2$ ), there are $2 \times 3^{l}$ proper prefixes $q$ of length $l+1$ ($0 \leq l+1 < k$) : 
\begin{enumerate}
\item the first symbol must be ${\tt 1}$, or $\underset{\bullet}{\tt 1}$ (two possibilities),
\item the next $l$ symbols can be ${\tt 0}$, ${\tt 1}$ or $\underset{\bullet}{\tt 1}$ ($3^l$ possibilities).
\end{enumerate}

Adding the initial state, gives :

\[
\underbrace{1}_{\textrm{initial state}}
\quad + \quad 
\underbrace{\sum_{l=0}^{r} 2 \times 3^{l}}_{\textrm{proper prefixes of length $0 \leq l+1<k$}} 
\]
\[=\]
\[
1
\quad + \quad 
2 \times \frac{3^{r+1} -1 }{2}
\]
\[=\]
\[
3^{r+1}
\]
This bound is reached for the same reasons of non-reducibility (applied on transition labels on Mealy, and not on final state labels as in Moore).

\end{proof}

\subsection{Coverage automaton size in the general case}
\label{subsection:CoverageSizeGeneral}

Now consider a seed of span $k$ with $r$ jokers and of weight $w$ ($w+r=k$). Following the previous section \ref{subsection:CoverageSize101}, a similar reasoning gives a bound on the automaton size of $2^w \times 3^r$ for the Moore automaton and of $(2^w -1) \times 3^r$ for the Mealy automaton.

\begin{proof}\itshape
We concentrate first on the Moore automaton.
We respectively call $r_j$ and $w_j$ the number of {\em joker} symbols and {\em must match} symbols for a given seed prefix of length $j$ ($r_j = \big|{}_j(\pi)\big|_{*}$, $w_j = \big|{}_j(\pi)\big|_{1}$, $r_j + w_j = j$). 
We don't necessarily suppose that the seed starts and ends with a {\em must match} symbol.
We will show that the number of states $\langle\substack{q^{} \\ c^{}}\rangle $ such that $|q|$ and $|c|$ are $\leq j$ is at most $2^{w_j} \times 3^{r_j}$, by induction.

\begin{itemize}
\item This is first true for $j = 0$, because the {\em empty state} (also called the {\em initial state}) $\langle\substack{q^{} \\ c^{}}\rangle$ where $|q|=|c| = 0$ is the only one that can match the empty seed prefix ${}_0(\pi)$.

\item If we suppose that it is true for a given $i$ ($\#\{states \quad \langle\substack{q^{} \\ c^{}}\rangle \quad\mathrm{with}\quad |q|=|c| \leq i\} \leq 2^{w_i} \times 3^{r_i}$), it can be now considered for $j=i+1$. We split the demonstration for $j$ in two parts :
\begin{enumerate}
  \item when $|q|=|c| \leq i$, by taking the set of the $2^{w_i} \times 3^{r_i}$ possible $\langle\substack{q^{} \\ c^{}}\rangle$ states (induction hypothesis)
  \item otherwise, when $|q|=|c| = j$, by considering and adding to this set the states $\langle\substack{q^{} \\ c^{}}\rangle$ that can be possibly reached. Two cases must then be considered : 
  \begin{enumerate}
    \item[(a)] if the last symbol $\pi[j]$ of the seed prefix ${}_j(\pi)$ is a {\em must match}, this symbol can only be compatible with a {\tt 1} on $q[j]$ (and this {\tt 1} cannot be covered by $c[j]$, as the last one being added). 
    \item[(b)] if the last symbol $\pi[j]$ of the seed prefix ${}_j(\pi)$ is a {\em joker}, this symbol can be compatible either with a {\tt 0} or a {\tt 1} on $q[j]$ (which cannot be covered by $c[j]$ too) . 
  \end{enumerate}

  Considering now the prefix ${}_i(\pi)$ preceding $\pi[j]$, we can see that : 
  \begin{itemize}
    \item the $w_i$ {\em must match} symbols of ${}_i(\pi)$ are compatible with a ${\tt 1}$ or a $\underset{\bullet}{\tt 1}$ ($2^{w_i}$ possibilities), 
    \item the remaining $r_i$ {\em jokers} of ${}_i(\pi)$ are compatible with a ${\tt 0} $, a ${\tt 1}$ or a $\underset{\bullet}{\tt 1}$ ($3^{r_i}$ possibilities). 
  \end{itemize}

  Combining each of the cases (a) and (b) with the preceding prefix ${}_i(\pi)$ gives $1 \times 2^{w_i}3^{r_i}$ states for (a), or $2 \times 2^{w_i}3^{r_i} $ states for (b), respectively, when $|q|=|c| = j$.

\end{enumerate}
At the end, because (a) $w_j = w_i+1$ and $r_j = r_i$, or (b) $w_j = w_i$ and $r_j = r_i+1$ otherwise, we can see that summing the number of states when $|q|=|c|\leq i$ and when $|q|=|c| = j$ gives the expected result $2^{w_j} \times 3^{r_j}$, for {\em non-final states}. 

It must be then noticed that, even for {\em final states}, new symbols that have just been covered ($\underset{\circ}{\tt 1}$) are only replacing the non-covered ones ({\tt 1}) on a subset of the $w$ fully determined positions given by the seed shape that are not yet covered ($\underset{\bullet}{\tt 1}$): they thus don't modify the recurrence when $j=|\pi|$ as they simply represent indicators to compute the coverage increment.
\end{itemize}

\medskip

We concentrate now on the Mealy automaton. Again, the main difference with the Moore automaton is that suffixes of full length $k$ are not represented because coverage values are set on transitions, but one thing to consider is that the last symbol can be covered on the Mealy automaton \citep[see][]{MartinNoeUn14}. By a similar reasoning, there are thus at most $\sum_{i=0}^{k-1=r+w-1} 3^{r_i} \times 2^{w_i} $ states in the general case, and :

\[
\sum_{i=0}^{r+w-1} 3^{r_i} \times 2^{w_i} \leq \sum_{i=0}^{r} 3^i + 3^r \sum_{i=1}^{w-1} 2^i = 2^w 3^r - \frac{3^r+1}{2} < 2^w 3^r
\]

Note that if we suppose that the seed starts with a {\em must match} symbol, then this bound can be reduced a little more :

\[
\sum_{i=0}^{r+w-1} 3^{r_i} \times 2^{w_i} \leq 1 + 2 (\sum_{i=0}^{r} 3^i + 3^r \sum_{i=1}^{w-2} 2^i ) = (2^w -1) 3^r
\]

\end{proof}

We notice in practice a much smaller size, and we suspect this bound more likely to be a $polynom(w,r) \times 3^r$ value, instead of exponential both in $2^w$ and $3^r$. In the special case of symmetric seeds, we already have a very simple proof of this $polynom(w,r) \times 3^r$ bound. This is interesting, because experimentally, seeds of the form {\tt 11${}^u$(*1${}^u$)${}^r$1} have been shown to give large coverage automata size.

Note even if not satisfying, the general result still improves on the only available ``bound'' proposed to date \citep[in][]{BensonMakSPIRE08} that can be estimated to be of order $w 2^w 4^r$.

\end{document}